\documentclass[12pt]{article}
\usepackage{geometry}
\usepackage[T1]{fontenc}
\usepackage[utf8]{inputenc}
\usepackage{lmodern}
\usepackage{amsmath,amssymb}
\usepackage{graphicx}
\usepackage{booktabs}
\usepackage{csquotes}
\usepackage{array}
\usepackage{caption}
\usepackage{float}
\usepackage{xcolor}
\usepackage{setspace}
\usepackage{enumitem}
\usepackage{natbib}
\defcitealias{cftc2026compute}{CFTC, 2026}
\usepackage{hyperref}
\hypersetup{colorlinks=true,linkcolor=blue,citecolor=blue,urlcolor=blue}

\title{\vspace*{-0.5in}(Early) AI Compute Asset Pricing\thanks{We gratefully acknowledge Silicon Data and ORNN for providing data. Bandi: \href{mailto:fbandi1@jhu.edu}{\texttt{fbandi1@jhu.edu}}; Su: \href{mailto:ys@jhu.edu}{\texttt{ys@jhu.edu}}. The newest version is available \href{https://www.suyinan.com/compute-ap-paper-redirect}{here}.}\\[0.5em]
}
\author{
\begin{tabular}[t]{cc}
Federico M. Bandi ~~~~~& Yinan Su
\\
\small Johns Hopkins University ~~~~~ & \small Johns Hopkins University
\end{tabular}\\[0.55em]
}
\vspace{1cm}
\begin{document}

\date{\small This draft: October 20, 2026~~~~~First draft: June 30, 2026}

\maketitle

\begin{abstract}
Compute (computing power) is a scarce, capital-intensive  input at the center of the AI economy. Compute capital expenditure and service flow already exceed 1\% of U.S.\ GDP and are growing rapidly. The price of compute reflects uncertainty over AI adoption. The announced launch of \textit{compute futures} turns this uncertainty into a tradable risk, raising questions on the pricing of a new asset class. We provide an early asset-pricing framework for compute. We begin by discussing the underlying compute rental market and its indexation. We then turn to pricing: 1) direct no-arbitrage links between futures prices and current spot prices fail due to the non-storable nature of compute, 2) \textit{synthetic} futures prices from existing term rental contracts are likely upper bounds on true futures prices and, 3) upon financialization, futures prices will be investors' expectations of spot prices at expiration net of a risk premium. Using \textit{synthetic} futures as stand-ins before the compute futures market launches, we construct the first compute futures return panel sorted by GPU generation and maturity. Our preliminary evidence is consistent with a \textit{positive} compute risk premium, suggesting hedging pressure on the part of compute providers.

\end{abstract}

\begin{flushleft}

\end{flushleft}

\clearpage
\section{Introduction}

\begin{center}
\small\emph{``As the backbone of the digital economy, compute is the new oil of the 21st century.''}
\end{center}
\vspace{-1.5em}
\begin{flushright}
\small --- Terry Duffy, Chairman and CEO of CME Group\footnote{``CME Group and Silicon Data Partner to Launch First Compute Futures,'' May 12, 2026, \href{https://www.cmegroup.com/media-room/press-releases/2026/5/12/cme_group_and_silicondatapartnertolaunchfirstcomputefutures.html}{link}.}
\end{flushright}

\vspace{-0.2em}

\begin{center}
\small\emph{``I actually believe a new asset class will be buying futures of compute.''}
\end{center}
\vspace{-1.5em}
\begin{flushright}
\small --- Laurence D. Fink, Co-founder, Chairman, and CEO of BlackRock\footnote{``A Conversation with BlackRock CEO Larry Fink and Brookfield Corporation CEO Bruce Flatt,'' Milken Institute Global Conference transcript, May 2026, \href{https://milkeninstitute.org/sites/default/files/2026-05/AConversationBlackRockCEOLarryFinkBrookfieldCorporationCEOBruceFlatt_Transcript_GC26.pdf}{link}.}
\end{flushright}

Compute (computing power) is a central commodity of the AI economy. It is a scarce, capital-intensive input that powers frontier model training, large-scale inference and, more generally, the delivery of AI services. The current price of compute reflects, e.g., the cost of producing AI itself. Future compute prices embed uncertainty over the scale of AI adoption.

On May 12, 2026, the CME Group announced plans, pending regulatory review, to launch a \enquote{first-in-class \textit{compute futures market} later in 2026} with contracts designed to help typical compute market participants (AI developers and compute-service providers) and beyond (financial institutions) manage compute price risk.\footnote{\label{fn:cme-silicon-launch}According to the announcement, the products will be based on Silicon Data's daily GPU benchmarks for \texttt{on-demand} rental rates and are designed to ``allow traders, financial institutions, AI builders and cloud-service providers to manage volatility and price risk associated with the multi-trillion-dollar compute market.'' ``CME Group and Silicon Data Partner to Launch First Compute Futures,'' May 12, 2026, \href{https://www.cmegroup.com/media-room/press-releases/2026/5/12/cme_group_and_silicondatapartnertolaunchfirstcomputefutures.html}{link}.}\textsuperscript{,}%
\footnote{\label{fn:ice-ornn-launch}Within a week, ICE and Ornn announced a similar plan, subject to regulatory approval, to launch U.S. dollar-denominated, cash-settled GPU compute futures based on Ornn's Compute Price Index (OCPI), a transaction-based GPU pricing benchmark. ``ICE and Ornn to Launch GPU Compute Futures Contracts,'' May 19, 2026, \href{https://ir.theice.com/press/news-details/2026/ICE-and-Ornn-to-Launch-GPU-Compute-Futures-Contracts/default.aspx}{link}. Both Silicon Data and Ornn are start-up companies providing compute market data and indexation. This article analyzes data from both providers, as we detail further below.}
The impending arrival of a compute futures market will make compute tradable, thereby creating a financial instrument for hedging and price discovery, i.e., one that will assist the risk management and financing of the AI infrastructure.
We ask how this new asset class should be priced - and what the limited data currently available suggests about implications from theory - \textit{before} the compute futures market is formally launched.

Our approach in this article is not dissimilar from what is routinely done in other fields. As a (admittedly hyperbolic) example, how should a new helicopter fly on Mars, with no opportunity of a real test flight before a multibillion-dollar launch and a seven-month journey? NASA's ingenuity addressed the question by adapting the physics and engineering principles learned from flights on Earth to the new environment of lower gravity and thinner atmosphere.\footnote{Mars has only about one-third of Earth's gravity but roughly one percent of its atmosphere. Thus, the same physics of lift had to be applied under radically different parameters, including much larger rotor blades and much higher rotation speeds compared to Earth helicopters. The design process involved careful modeling and simulation, but no real-world test flights until the helicopter arrived on Mars.}
We apply the same approach to the emerging class of AI compute assets. We ask which established financial principles carry over to AI compute asset pricing and how they should be adapted to the new environment.

The economic importance of this problem stems from the position of compute in the AI economy. NVIDIA CEO Jensen Huang has made famous a description of the AI stack as a five-layer cake:
\[
\text{Energy} \Rightarrow \text{Chips} \Rightarrow \text{Infrastructure} \Rightarrow \text{Models} \Rightarrow \text{Applications}
\]
The compute market sits at the interface between infrastructure and model production. Its price, therefore, contains information about both sides of the AI economy. On the supply side, it reflects the cost of deploying and operating data centers, which in turn depends on the cost of chips, energy and other inputs. On the demand side, it reflects the value of training and inference workloads, which in turn depends on the productivity of AI applications and, ultimately, on the scale of AI adoption. Should AI become a general-purpose technology, compute prices will not merely be input costs of one economic sector. They will be the prices for the one binding resource behind the broad AI economy.

The economic scale of AI compute is already macroeconomically material. Based on our calculations, the 2025-Q4 installed compute stock already implies a gross compute service flow of around
\$430 billion to \$1.3 trillion per year, or approximately 1.4\% to 4.0\% of U.S.\ GDP.
In terms of investment, the planned 2026 capital expenditure by the four largest hyper-scalers (Amazon, Microsoft, Alphabet and Meta) is more than \$700 billion, or about 2.2\% of U.S.\ GDP.\footnote{See Appendix~\ref{app:compute-scale-calculations} for our sources, as well as simple calculations, behind these scale estimates.}
These magnitudes are already of macroeconomic significance for aggregate investment and output.
They may become substantially larger if current growth rates persist, but they are also subject to considerable uncertainty over model progress, chip supply, data-center deployment, energy constraints and AI adoption. The combination of uncertainly and scale is precisely what makes compute price risk an economically important object for trading and pricing.

The financialization of compute is consequential because the physical (spot) market alone constitutes an inefficient allocation mechanism for this risk. AI developers, cloud providers, data-center investors, chip producers, and financial investors all have exposure to future compute scarcity, but the current market mostly expresses that exposure through physical capacity contracts, bilateral cloud commitments, and equity prices. A compute futures market would separate price risk from physical delivery and operations. It would let users hedge future rental costs. It would let providers and infrastructure investors hedge revenue risk. It would also let outside investors take views on the future scale of AI demand, thereby generating a forward curve that can discipline capacity investment and financing decisions. Once such a market exists, compute becomes an asset-pricing object: futures prices, forward curves, expected returns, return volatility and risk premia become central quantities for understanding how the market prices the future of AI.

\begin{figure}[!t]
\centering
\vspace{-0.8em}
\includegraphics[width=0.72\linewidth]{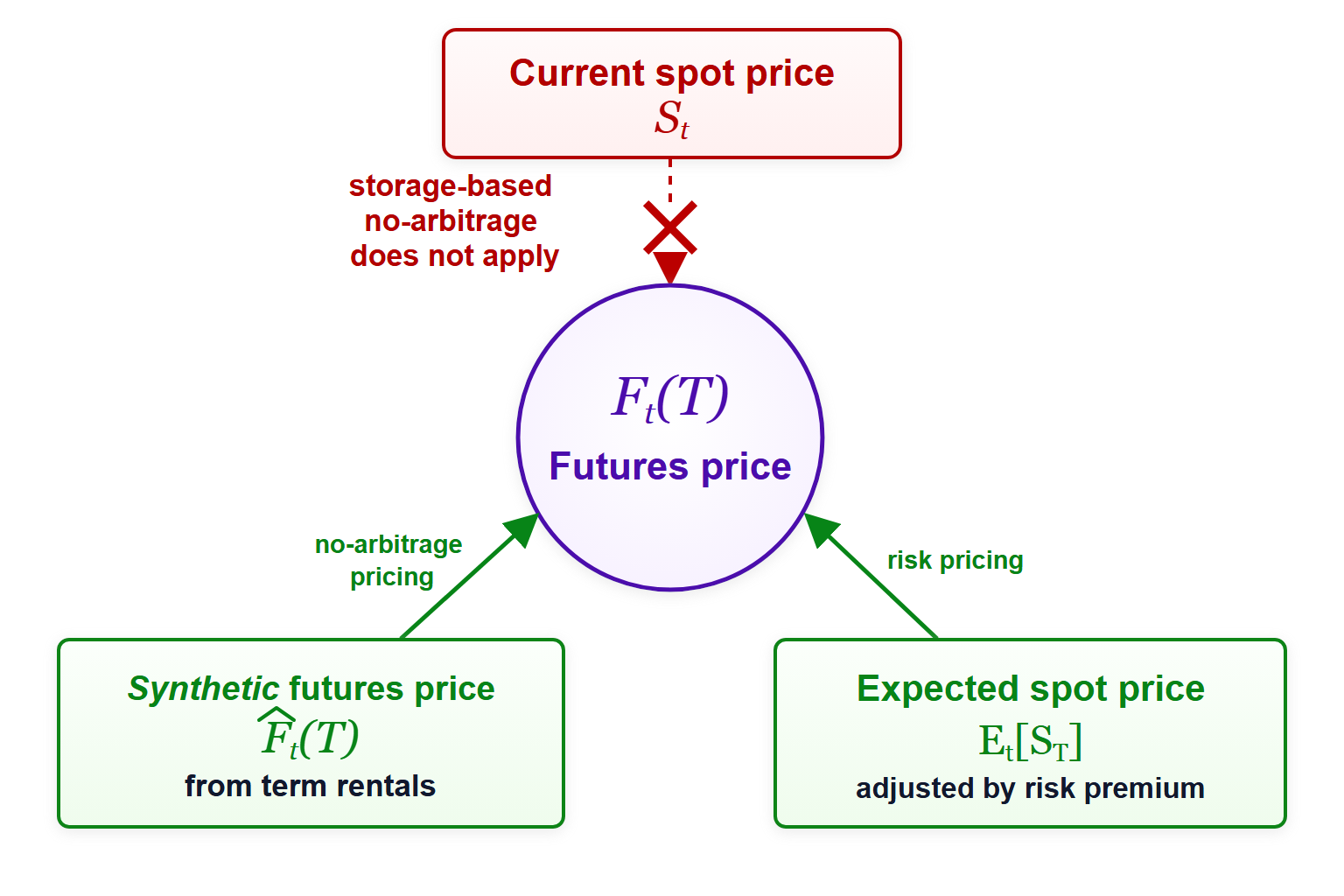}
\caption{\textbf{The pricing of compute futures.} The dashed upper spoke indicates that the standard storage-based arbitrage link from current spot prices to futures prices would not apply to compute rental service. The solid lower spokes indicate the two pricing approaches discussed in the paper: no-arbitrage pricing against term rental forward contracts and no-arbitrage risk pricing against expected spot prices at delivery.}\label{fig:pricing-links}
\end{figure}

We take a first step towards the asset pricing of AI compute. Compute is an emerging commodity for which we do not enjoy a history of traded futures prices. Thus, the goal cannot be to impose a mature theoretical and empirical framework. Rather, our objective is to identify which pricing restrictions should apply to compute futures and which familiar restrictions should fail (c.f. Fig.~\ref{fig:pricing-links}). We then discuss how the existing term-rental compute market can be used to construct \textit{synthetic} futures prices and what a first panel of \textit{synthetic} futures prices currently implies about the risk and return of this new asset class. We do so by utilizing term-rental data made available to us by Silicon Data, a data provider specializing in compute market data and the very company partnering with the CME to launch the first compute futures contract.

The theoretical analysis makes three points. First, the (storage-based) no-arbitrage logic to derive forward/futures prices from current spot prices for storable commodities should not directly apply to compute. A GPU-hour not used today cannot be carried forward and delivered in the future. Thus, there is no direct cash-and-carry link from the current spot rental price, \(S_t\), to the futures price, \(F_t(T)\). In this sense, compute is closer to electricity and other capacity-constrained service commodities than, say, to oil. The relevant economic link is, therefore, not between today's spot rental price and the price for future delivery, but between the price for future delivery and the expected spot rental price at delivery.

Second, although storage-based no-arbitrage arguments fail, existing term-rental contracts provide a useful no-arbitrage reference point. A contract that locks in compute capacity over a future rental interval can be transformed into a \textit{synthetic} futures contract. Ignoring marking-to-market in the futures market, the traded futures price \(F_t(T)\) should be close to the \textit{synthetic} futures price inferred from the term rental curve. We argue that, in the current market, the \textit{synthetic} futures price is likely to be an upper bound on the financial futures price. Physical term rentals bundle price insurance with a real capacity-locking option, and that option is valuable when users have limited alternatives for securing scarce compute. This suggests that \textit{synthetic} futures are likely more expensive than financial futures, with the wedge most visible at short maturities as the capacity-locking premium runs off toward spot delivery.

Third, compute futures should carry a risk premium. The core risk-return relation links \(F_t(T)\) to the conditional expectation of the spot rental price at delivery, i.e., \(S_T\):
\[
F_t(T)=\mathbb{E}_t[S_T]-\lambda_t(T).
\]
The sign of \(\lambda_t(T)\) depends on how compute prices covary with the marginal utility of the \enquote{marginal} investor. Upon launch, natural hedgers focused on operational risk (like AI developers and compute-service providers) are likely to drive the market. Should compute providers be the \enquote{marginal} investor, the covariance would likely be negative and the compute risk premium would likely be positive. In other words, the long positions would be \enquote{offering} insurance to the providers and a positive compute risk premium would be compensation for the provision of insurance. The outcome would be futures prices lower than the expected compute price at delivery.\footnote{The expected spot rental price at delivery is a particularly hard-to-pin-down quantity, as the future of AI and compute demand is the subject of a somewhat heated debate. The pessimistic view that the current massive compute investment wave is an unwarranted bubble - and that AI demand will ultimately disappoint - would result in a low expected spot rental price. The optimistic view suggests, instead, that AI demand will accelerate as models become more capable and applications proliferate. In this scenario, physical supply may find it hard to keep up with growing demand, which would result in a high expected spot rental price. The return volatility in the semiconductor and cloud provider sectors in late 2025 to early 2026 is evidence of uncertainty and disagreement in the market. A compute futures market will allow uncertainty and disagreement to be quantified directly. This article does not attempt to predict the expected spot rental price, although the economics of the problem is of interest and will be a topic for future work.}

Because a history of traded compute futures does not yet exist, we build the first \textit{synthetic} futures return panel from Silicon Data rental curves. The Silicon Data rental curves are particularly informative in that the CME is expected to launch futures contracts on Silicon Data indices. Our panel is a two-dimensional cross section. One dimension is compute vintage, represented by A100, H100, and B200. The other is time to maturity. This is the compute-market analogue of an asset-pricing return panel organized by economically meaningful sorts. We construct held-to-maturity futures returns and, consistently with the view of a market currently dominated by compute providers and their hedging decisions, estimate (largely) \textit{positive} risk premia. We also construct rolling constant-maturity futures returns and the corresponding cumulative return series. As for more traditional asset classes, the cumulative return series show persistent co-movement across maturities and visible maturity spreads. We document when our proposed \textit{synthetic} futures construction is informative and when it is affected by maturity run-offs and early data quality.

A natural objection is that compute futures do not yet exist and compute forward data is limited for a complete asset-pricing treatment. We view the centrality of compute in today's AI economy as a reason to rationalize the market early rather than a reason to ignore it. \citet{akerlof2020sins} argues that economics can commit ``sins of omission'' when the profession favors problems that can be treated with established hard methods and neglects timely problems whose appropriate tools are still developing. His conclusion is methodological rather than anti-rigorous: ``Different terrains call for different vehicles.'' Compute futures may be one such terrain. 

The simplicity of our study is, therefore, intentional. In a new market, the first contribution is to define the object, identify the applicable pricing restrictions, document the available data and separate robust implications from hypotheses that require a trading history. We do so below.

\section{Related literature}

To the best of our knowledge, this article is the first to study compute as an emerging asset class and to provide a theoretical and empirical analysis---albeit an early one---of (aspects of) compute asset pricing.

Because of the growing scale and economic significance of AI adoption, a recent literature has, however, begun to examine certain financial features of the AI infrastructure:
\citet{wachter2026investment} are interested in what investment data imply about the AI transition and
\citet{vannieuwerburgh2026financing} discusses the financing of the AI buildout and the allocation of infrastructure risk. 
Compute is an output of the AI infrastructure. The financialization of compute---our focus---is expected to facilitate the investment in, and the financing of, the AI infrastructure via price discovery and risk transfer. In this sense, the asset pricing of compute should not be viewed as a narrow question concerning a nascent class of financial instruments, but rather as an open field of study on the valuation of a central element of the AI economy. 

Downstream to the compute market, \citet{demirer2025emerging} study the supply, demand and pricing of large language model inference services, while \citet{timmermann2026compute} highlight the role of compute as a scarce capital resource with the potential to improve performance in return prediction and trading. Related to our discussion (below) of AI compute indexation, \citet{byrne2021cloud} construct hedonic price indexes for Amazon Web Services in the context of cloud computing. \citet{bergemann2025robust} study the optimal contract of a cloud provider facing uncertainty about demand.

More broadly, this article is also related to the literature that examines the economic implications of AI. \citet{brynjolfsson2017paradox}, \citet{acemoglu2024simple} and \citet{korinek2024concentrating}, e.g., offer early insights into the industrial organization and macroeconomics of AI adoption.

In asset pricing, we draw from the literature on commodity futures, in particular from work on electricity futures (\citealp{bessembinder2002equilibrium}, and \citealp{longstaff2004electricity}) and on risk premia and hedging in commodity futures markets (\citealp{deroon2000hedging}, and \citealp{szymanowska2014anatomy}). \citet{fama1987commodity} is a classic reference on the economics of commodity futures. More recently, \citet{he2025fundamentals} have focused on a new futures product, \enquote{perpetual} future, and have studied its pricing. In principle, perpetual futures may assist the financialization of compute.

We note that recent practitioner reports and commentaries on compute as a commodity and its financialization have been issued, e.g., by \citet{friedman2026compute} and \citet{belt2026commodity}.

\section{The underlying: compute market and its indexation}

This section reviews the current compute term rental market structure and the indexes that are foundational for structuring and pricing prospective compute futures contracts.

\subsection{The compute market}

The cash market underlying a prospective compute futures contract is the market for rented compute capacity. The relevant economic object is not ownership of a chip, but access to a standardized compute service bundle: accelerator time, host CPU and memory, storage, cooling, networking, location, reliability, and provider-specific operating quality. A futures contract on compute would, therefore, reference a benchmark rental price for this service flow. The conventional unit of rental price is dollars per GPU-hour or dollars per server/instance-hour.

The rental vs. ownership distinction matters for futures pricing. Installed computing power (GPU servers and data centers) is a capital good that can be owned, depreciated, sold, and redeployed. Rented compute capacity is, instead, a service flow. A GPU-hour that is unused today cannot be stored and delivered next month. In this respect, compute is closer to electricity, freight, or other capacity-constrained service commodities than to a storable physical commodity, such as oil or metals. The relevant spot price is the price of access to capacity at a point in time and the relevant forward price is the price of guaranteed access to capacity over a future interval. Standard cash-and-carry intuition for pricing compute forward would, therefore, hardly apply. We will return to this point.

Compute rental prices are economically important because they reflect, among other issues, the cost of access to a scarce input in AI production. Frontier model training, inference at scale, and enterprise AI applications require not only GPUs, but GPUs embedded in data centers with power, cooling, memory, storage, interconnect, and operational reliability. The rental price of GPU capacity, therefore, reflects both current demand for AI workloads and the short-run scarcity of deployable infrastructure.
Looking forward, the path of compute rental prices will depend on the future trajectory of AI: whether new models and applications make AI broadly more valuable, whether enterprise adoption turns capability gains into sustained compute demand, and whether supply-side bottlenecks in chips, data centers, power, and deployment ease or intensify. In essence, futures contracts written on compute will be a natural way for financial markets to aggregate and reveal information about the future of the AI economy.

The compute term rental market has heterogeneous products, providers, and rental terms. It, therefore, requires careful treatment before its prices can be used as the underlying in futures contracts. The problem is familiar from, e.g., the automobile and the housing markets. Regarding the former, a price is only meaningful after specifying the bundle being priced, because model, mileage, and other characteristics can make superficially similar units economically distinct.

First, the typical service differentiator is the GPU benchmark, the most relevant benchmarks for AI workloads in the current market being NVIDIA A100, H100, B200, AMD MI300X, etc. However, the rented object is not an isolated GPU, although the price unit convention is dollars per GPU-hour. Rather, the purchased service is capacity from a server- or rack-scale system that bundles GPUs with CPUs, memory, networking, storage, power, cooling, software, and operational management. NVIDIA CEO Jensen Huang emphasized the industry shift toward system-level compute in his GTC 2026 keynote:

\begin{quote}
``\textit{In the past, when I mentioned Hopper, I would hold up a chip, which was quite cute. But when I mention Vera Rubin, people think of the entire system.}''\footnote{Hopper refers to the architecture underlying the H100 GPUs in our data, while Vera Rubin refers to NVIDIA's next-generation rack-scale AI computing platform. }
\end{quote}

The implication is that the observed price of a GPU-hour reflects not only the accelerator itself, but also the surrounding infrastructure required to make the accelerator productive.
Thus, constructing a tradable compute index requires adjustments for differences across hardware generations, server configurations, interconnect topology, memory capacity, reliability guarantees, and associated service characteristics. A raw average of quoted GPU-hour prices would otherwise combine economically distinct compute products and introduce substantial measurement error.

Second, the current compute market has two broad provider categories: hyper-scalers and neo-clouds. 
Examples of hyper-scalers include AWS, Microsoft Azure and Google Cloud. Examples of neo-cloud providers include CoreWeave, Lambda, Crusoe, Nebius and many others. Hyper-scalers are vertically integrated cloud platforms that supply compute alongside storage, networking, enterprise software, and other complementary services.
They account for the larger share of the market. 
Their compute capital is financed through large corporate balance sheets and diversified cash flows from more traditional tech services. 
Neo-clouds are smaller, but faster-growing, specialists. Because they are more narrowly exposed to AI compute, their expansion has often relied on private equity and credit, customer contracts and GPU-backed debt.
In practice, hyper-scaler rental prices are typically higher - and more stable over time - than comparable neo-cloud prices. The premium reflects a broader cloud compute bundle: reliability, compliance, geographic redundancy, enterprise support, and ecosystem lock-in. 
Neo-cloud prices are often closer to the marginal rental cost of GPU capacity, making them informative about the short-term supply and demand fluctuations in the compute market.

Third, there are different rental contract forms. \texttt{On-demand} rentals give flexible access to capacity at the current date, without a long-term commitment. The corresponding price is the closest analogue to a financial spot price. Another type of rental contract, called \texttt{spot}, offers access to excess capacity at a lower price, but with the risk that the provider may reclaim the resource when demand rises. We note that \enquote{spot} in cloud-provider terminology is different from \enquote{spot} in financial-derivatives terminology. \texttt{Reserved} rentals, by contrast, entail an advance commitment 
to use capacity over a specified future period, typically 1 year to 3 years. These contracts, therefore, bundle a forward commitment for capacity reservation and, sometimes, discounts for various levels of reduced flexibility. The \texttt{Reserved} rental price provides information about the forward price of compute, from which \textit{synthetic} futures prices can be constructed. In this sense, \texttt{Reserved} rental data are a critical input for deriving guidelines for the prospective compute futures market without an history of actual futures trades. Our empirical analysis will rely on available data on the \texttt{Reserved} rental segment of the compute market.

Next, we present a couple of concrete examples of rental listings posted some time in May 2026 to illustrate the structure and primitives of typical rental contracts. 

\begin{quote}
AWS: \texttt{on-demand} instance \texttt{p5.48xlarge} (AWS name for an H100 server) \texttt{Linux} in \texttt{us-east-1} (N. Virginia), priced at \$55.04 per instance-hour, bundling 8 H100 GPUs, 640 GB HBM3 GPU memory, 192 vCPUs, 2,048 GiB system memory, and 3,200 Gigabit networking. Translating to \texttt{6.88 dollars per GPU-hour}.
\end{quote}

\begin{quote}
CoreWeave: \texttt{on-demand} instance \texttt{NVIDIA HGX B200} at \$68.80 per hour. The listed bundle contains 8 B200 GPUs, 180 GB of VRAM per GPU, 128 vCPUs, 2,048 GB of system RAM, and 61.44 TB of local storage. It translates to \$8.60 per GPU-hour. A similar \texttt{HGX B200} instance with \texttt{spot} rental for \$34.11 per hour, or \$4.26 per GPU-hour.
\end{quote}

Notwithstanding the heterogeneity of the compute market, a certain degree of substitutability exists.
In particular, the emergence of compute capacity intermediaries facilitates substitutability. 
One example of such a mechanism is NVIDIA DGX Cloud Lepton, an AI compute marketplace and platform for finding and using GPU capacity across multiple cloud providers. At the model level, routing platforms such as OpenRouter provide a single API endpoint for accessing hundreds of LLMs from many providers, without requiring users to interact separately with OpenAI, Anthropic, or open-source model providers.
When workloads are containerized and run on common software stacks, users can increasingly submit a job to an intermediate platform that searches across providers, regions, and GPU types for available capacity. 
These platforms reduce switching costs across clouds and make capacity from different providers closer substitutes. 
Economically, these routing layers act like compute intermediaries: they aggregate fragmented GPU supply and transform heterogeneous provider-specific offerings into a more connected market for compute capacity.

Substitutability implies common fluctuations in rental prices across different service configurations, providers and rental forms. Substitutability, and the implied common price fluctuations, is the premise for the construction of standardized compute price indexes. This is analogous to how stock market indexes, such as the S\&P 500 index, capture common factors in the cross-section of stocks. A potential factor structure is also the premise for computer providers and users whose cash market exposure is limited to a segment of the compute market to use the prospective compute futures as a hedging and risk management tool. Studying the factor structure of the compute market and the interpretation of alternative risk factors is, therefore, an important direction for future research, one that mature financialization will facilitate. Below, we present a preliminary analysis of substitutability and co-movement at the index level.

We have, thus far, only touched upon the basics of the compute market structure in order to motivate its indexation and financialization. The compute market, however, has many other features, from capital financing and investment to contract design and market power, that are of interest for economic analysis and policy making. In the end, the pricing of financial assets on compute will be linked in important ways to the industrial organization of the underlying compute market. 

\subsection{Compute price indexation}

Cash-settled compute futures require an index that summarizes the rental price of the underlying compute service. At least two start-up companies, Silicon Data and Ornn, have compiled and launched public compute price indexes based on prices from different providers. The current indexation takes a ``surveyed'' approach: the indexes are aggregation of observed heterogeneous compute prices. Both Silicon Data and Ornn have announced plans to launch compute futures contracts with, respectively, CME and ICE in which their indexes are the underlying (c.f. Footnotes \ref{fn:cme-silicon-launch} and \ref{fn:ice-ornn-launch}). The August 2026 request for comment (RFC) by the Commodity Futures Trading Commission (CFTC) on the listing of compute derivatives contracts, which cites an earlier version of this paper, explicitly asks whether compute price series can provide reliable settlement references and whether the contracts would be readily susceptible to manipulation \citepalias{cftc2026compute}.

A transparent, representative, and manipulation-proof index is a critical foundation for a successful futures market.

Transparency requires that users understand which providers and rental instances enter the index, how their prices are obtained, and how those observations are selected. The methodology, conversion factors, trimming rules, and input selection criteria should be publicly documented. The index should also be reproducible by market participants and auditors and rely on transactions that can be easily verified. The fact that compute transactions are largely OTC can be addressed through a well-functioning trade reporting infrastructure, with transactions recorded and anonymized for privacy. Market participants and auditing companies should be able to download the data and apply the published calculation, with published corrections for deviations from the reference compute service. Should liquidity in the OTC market not be high enough, the index administrator could rely on binding quotes published by a contributor panel of hyperscalers and neoclouds. Such quotes would need to be firm and executable for a specified quantity during the fixing window.

Representativeness requires that the underlying observations adequately capture the market segment the index is intended to measure. Compute markets are segmented across provider types, GPU models, and contractual arrangements. Privately negotiated, longer-term contracts in the hyperscaler segment may generate relatively infrequent price observations. The scope of a settlement index should therefore be distinguished from that of the broader compute rental market. For example, the proposed CME contracts reference Silicon Data's neocloud rental price indexes for specific GPU models---H100 and B200---measuring hourly on-demand rental costs. Even conditional on a GPU model, one GPU-hour can differ across providers in form factor, interconnectivity, and host and network specifications. One can adopt a ``par'' underlying that specifies these features: the underlying that would be delivered, if delivery could occur. If a recorded transaction does not use the ``par'' underlying, its price can be discounted or increased to capture deviations from the reference specification, using published quality adjustments. This is analogous to adjusting delivery prices for differences in grade in commodities such as wheat and corn. Hedonic modeling can inform the measurement of these differences, but its research objectives are distinct from the operational requirements of a futures settlement index.

Protection against manipulation requires safeguards for both the underlying observations and their aggregation. Existing futures contracts have addressed manipulation risk using volume-weighted trimmed means. A possible implementation for compute is a par-normalized volume-weighted trimmed mean: transaction prices are first adjusted for deviations from the reference specification, observations in the upper and lower tails are trimmed by volume, and the remaining prices are averaged using volume weights. Volume-weighted interpolated medians provide another approach to limiting the influence of extreme transaction prices. Neither method eliminates manipulation risk, and both require verification that the inputs reflect genuine arm's-length transactions. Independent audits, records of the underlying transactions and calculations, and controls over conflicts of interest among administrators and contributors would support these safeguards. Clear fallback procedures are also needed for periods when transaction volumes are insufficient or data feeds fail. These are requirements for assessing the suitability of an index for futures settlement, rather than an endorsement of the existing indexes used in our empirical analysis.

Appendix \ref{comments} discusses these issues further using a simple model. The Appendix contains our response to the CFTC's \enquote{Request for Comment on the Listing of Compute Derivatives Contracts,} published August 21, 2026, at 91 FR 54259 (RIN 3038-AF77; Docket No. CFTC-2026-1850).

\subsection{Index data and preliminary analysis}\label{index1}

We have access to six daily GPU rental index series from Silicon Data covering \texttt{on-demand} rental prices for A100 and H100 GPUs from both the hyper-scaler and the neo-cloud segment, as well as B200 and MI300X GPUs from the neo-cloud segment. Regarding Ornn, we have access to six daily GPU index series covering the neo-cloud segment. Table \ref{index} provides details.

\begin{table}[!t]
\centering
\resizebox{\linewidth}{!}{%
\begin{tabular}{lrllr}
\toprule
series (Bloomberg Ticker when available) & nobs & start & end of sample & latest (\$/GPU-hour) \\
\midrule
\textbf{Silicon Data} &  &  &  &  \\
A100 NEO \texttt{SDA100RT} & 595 & 2024-09-01 & 2026-04-18 & 1.430 \\
H100 NEO \texttt{SDH100RT} & 595 & 2024-09-01 & 2026-04-18 & 2.500 \\
B200 NEO \texttt{SDB200RT} & 261 & 2025-08-01 & 2026-04-18 & 5.100 \\
MI300X NEO & 108 & 2026-01-01 & 2026-04-18 & 2.380 \\
A100 HS & 595 & 2024-09-01 & 2026-04-18 & 3.690 \\
H100 HS & 595 & 2024-09-01 & 2026-04-18 & 7.430 \\
\textbf{Ornn} &  &  &  &  \\
A100 SXM4 & 835 & 2024-01-01 & 2026-04-14 & 1.073 \\
H100 SXM & 661 & 2024-06-23 & 2026-04-14 & 1.770 \\
H200 & 447 & 2025-01-23 & 2026-04-14 & 2.834 \\
B200 & 204 & 2025-09-23 & 2026-04-14 & 4.217 \\
RTX 5090 & 416 & 2025-02-23 & 2026-04-14 & 0.474 \\
RTX PRO 6000 WS & 211 & 2025-09-16 & 2026-04-14 & 1.034 \\
\bottomrule
\end{tabular}
}
\caption{\textbf{Daily compute price indexes.} In the Silicon Data's series, \texttt{HS} denotes \enquote{hyper-scaler} and \texttt{NEO} denotes \enquote{neo-cloud}. The Ornn data do not separately identify the hyper-scaler and the neo-cloud segments. However, they appear closer to neo-cloud data than to hyper-scaler data. \texttt{SDA100RT} etc. are the Bloomberg tickers for the Silicon Data rental indices. RTX 5090 and RTX PRO 6000 WS are workstation or prosumer-class GPU products rather than frontier cloud-compute GPUs. Thus, they are less central to our work.}\label{index}
\end{table}

Some indexes are publicly available. Silicon Data launched the H100 Rental Index on Bloomberg in May 2025 with ticker \texttt{SDH100RT} and the A100 and B200 Rental Indexes in December 2025 with tickers \texttt{SDA100RT} and \texttt{SDB200RT}. Ornn's Compute Price Index was announced as being available on Bloomberg in April 2026.

Next, we report basic descriptive statistics. Because of their relevance for the first generation of prospective compute futures contracts, we focus on the three Silicon Data neo-cloud benchmarks: A100 NEO, H100 NEO and B200 NEO. We first plot the daily time series. We then convert prices into H100-equivalent units. Finally, we describe the correlation structure across indexes.

\begin{figure}[!t]
\centering
\includegraphics[width=0.92\linewidth]{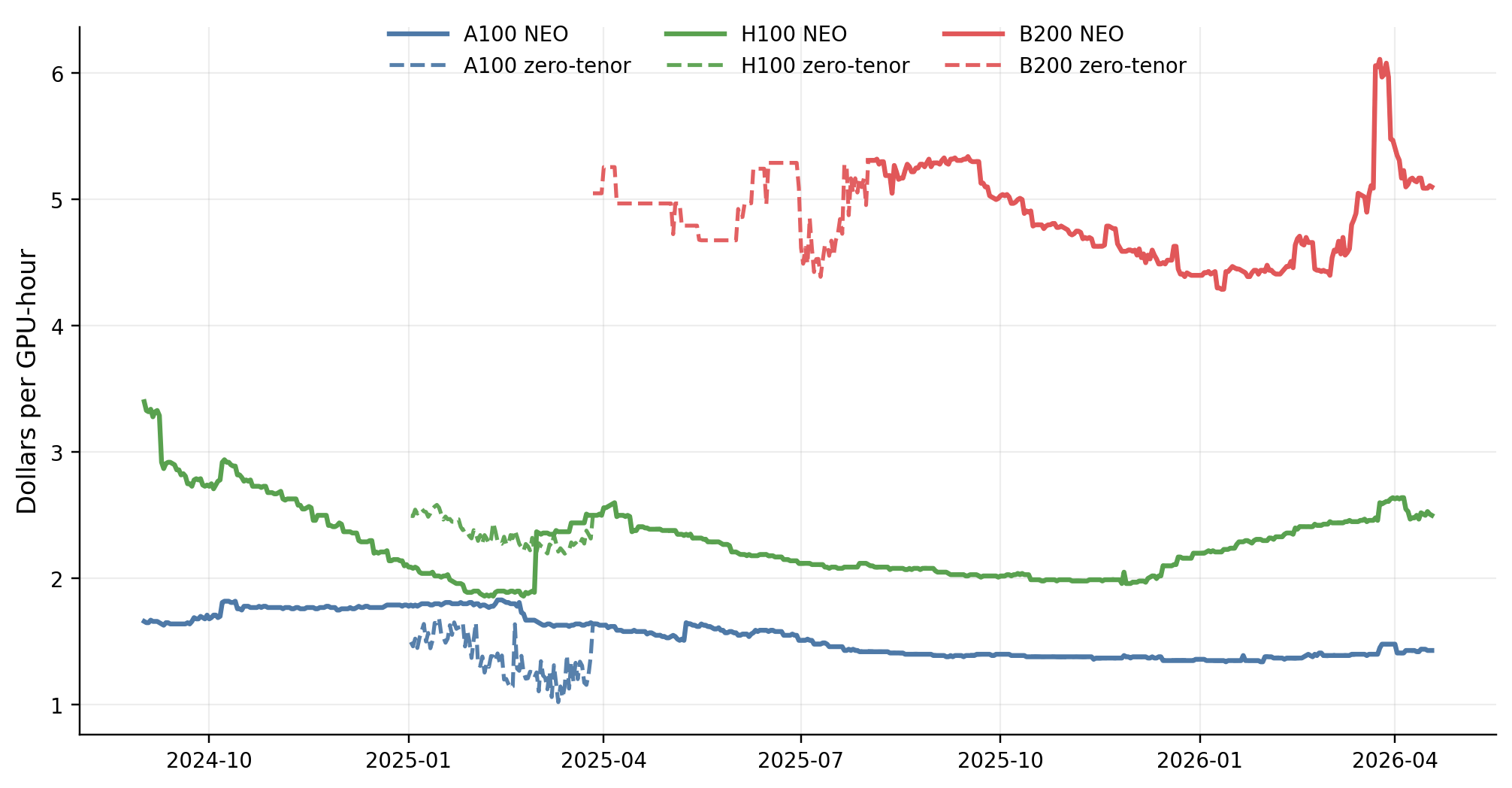}
\caption{\textbf{Compute price indexes (Silicon Data benchmarks, with zero-tenor \textit{synthentic} forward rates for verification.)} The solid lines are the three neo-cloud Silicon Data indexes. Dashed lines are the zero-tenor \texttt{term\_rate} observations from the Silicon Data forward curve for the same GPU family. The two series are closely aligned in the later part of the sample, but differ in the earlier part of the sample. This discrepancy suggests that the early forward-curve and spot-index records may not be fully harmonized. Empirical studies should use the later, cleaner part of the sample.}\label{fig:index}
\end{figure}

Fig. \ref{fig:index} shows, first, that newer and more capable GPU generations have higher rental prices per GPU-hour: A100 < H100 < B200. Below, we convert prices into H100-equivalent units to compare price per-unit of compute. 
Second, prices generally trend downward in the earlier part of the sample, while the H100 and B200 series turn upward beginning in late 2025 and into 2026. 
The downward pressure comes from the rapid expansion of compute supply, as new data centers and GPU servers are deployed. It is also the result of new chip generations being introduced. The upward pressure in late 2025 and into 2026 is likely driven by the surging demand for AI compute - as waves of agentic AI applications greatly increased inference demand - and the general acceleration of AI adoption.\footnote{OpenClaw, a self-hosted open-source AI agent, was first released in November 2025 and entered broader public discussion by early 2026, adding to the wave of agentic AI systems that shifted usage from simple chat interactions toward longer-running, tool-using workflows. OpenAI's codex, Anthropic's Claude Code, and other code-generation models also contributed to the surge in demand for AI compute in late 2025 and into 2026.}
A100 prices are less affected by the demand surge, possibly because the corresponding older-generation compute is more likely to be used for less demanding workloads and is, therefore, less sensitive to the recent waves of AI adoption. Third, we witness higher-frequency fluctuations with obvious positive correlations across the three indexes reflecting common short-run uncertainties of the compute market.

Next, we examine substitutability across GPU generations by converting prices into H100-equivalent units. The idea is to compare the rental price across GPU generations on a per-unit-of-compute basis.
If the production function of compute capacity \emph{scales} linearly with its compute power (measured by operations per second) then, with perfect substitutability, the price per unit of compute should be the same across GPU generations.

\begin{table}[!t]
\centering
\resizebox{\linewidth}{!}{%
\begin{tabular}{lrrr}
\toprule
 & A100 & H100 & B200 \\
\midrule
rental price per GPU-hour (end of sample) & \$1.433 & \$2.514 & \$5.111 \\
peak dense 8-bit compute per GPU, peta-operations per second (POPS) & 0.624 & 1.979 & 5.000 \\
H100-equivalent units per GPU & 0.315 & 1 & 2.527 \\
rental price per H100-equivalent-hour & \$4.543 & \$2.514 & \$2.023 \\
\bottomrule
\end{tabular}
}
\caption{\textbf{Rental prices converted to H100-equivalent units} The H100-equivalent compute unit and the conversion ratio (Peak dense 8-bit throughput) are from Epoch AI's GPU dataset.}\label{conversion}
\end{table}

The data shows that the price per unit of compute is close across GPU generations, but not equal, suggesting some degree of differentiation (c.f. Table \ref{conversion}). The price per H100-equivalent compute power is decreasing for newer and more powerful GPU generations:
A100 > H100 > B200.
This is expected given the very purpose of semiconductor scaling-ups: newer generations are designed to be more efficient and cost-effective. The older generation (A100) is almost twice as expensive on a per-unit-of-compute basis, reflecting its legacy demand for less demanding workloads. This is, again, analogous to the housing market: bigger houses often have lower prices per square foot, even though the total price is higher (adjusted for other quality characteristics),\footnote{In housing market hedonic models, the log-log coefficient of price on size is typically less than one, e.g.\ 0.707 in \citet{wooldridge2020introductory}.} suggesting sub-linear scaling of price with size.

Finally, we examine the correlation structure of the available indexes to further understand substitutability and co-movement across GPU generations and index providers.

\begin{figure}[!t]
\centering
\includegraphics[width=0.92\linewidth]{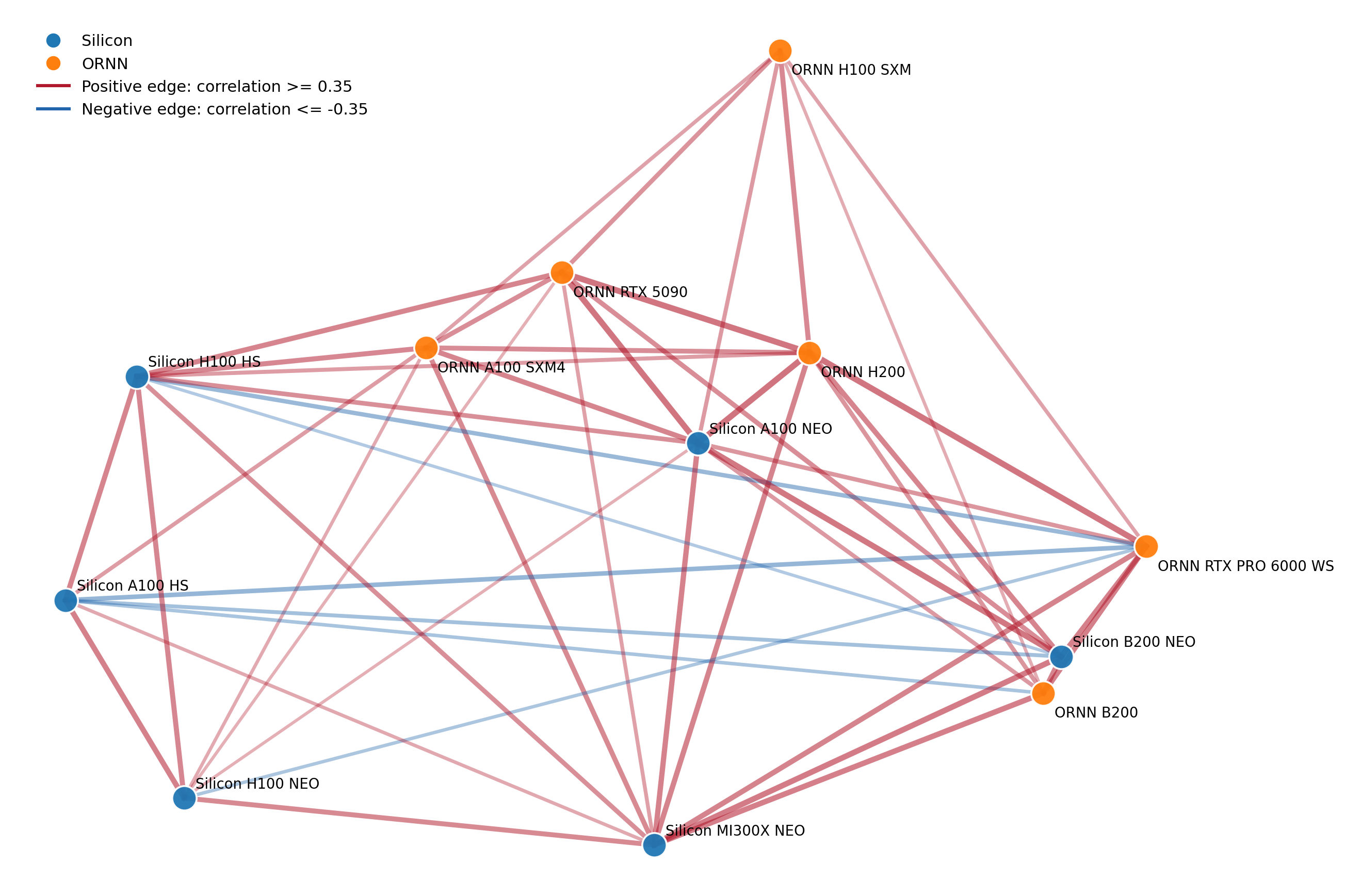}
\caption{\textbf{Correlation network across GPU rental price indexes.} The figure reports the correlation network for GPU rental price indexes from Silicon Data and Ornn. Correlations are computed on the trend component of each daily index, where the trend is a trailing 30-day exponential moving average of the raw index level. The recent 12-month sample is used when available. Otherwise, the figure uses the available overlap for the corresponding pair. Node positions are based on the correlation distance \(\sqrt{2(1-\rho)}\). Positive edges are drawn when the correlation is at least 0.35. Negative edges are drawn when the correlation is at most -0.35.}\label{network}
\end{figure}

The network visualization in Fig. \ref{network} shows both common and idiosyncratic movements. Most edges are characterized by positive correlations. Negative correlations, mostly across index provider, also exist. The Silicon Data indexes (blue) tend to be separated from the Ornn indexes (orange). This separation reflects either differences in index construction or differences in the market segments covered by the two providers (or a combination of both). Silicon Data A100 NEO is in a relatively central location. H100 NEO and B200 NEO are also connected to other indexes, suggesting relevance. On the left of the network, the two hyper-scaler indexes are close to each other, indicating that the hyper-scaler segment has its own common features.  

The disconnect between Silicon Data indexes and Ornn indexes is a potential issue. For instance, excluding the possibility of cross-hedging, a futures contract that references the Silicon Data's H100 NEO index is not an ideal hedging instrument for a user in the market segment covered by a negatively-correlated Ornn's index.\footnote{One example is the RTX PRO 6000 WS index which is in the market segment of workstation-level GPU. This segment is rather different from the higher-end data-center-level segment represented by H100 NEO.} Additionally, the forward-looking correlation structure relevant for hedging might be different from the short-term realized correlation estimated above from historical spot data.

\section{The financialization of AI compute}

Throughout the rest of the article, we use \(t\) and \(T\) to denote generic points in time, namely the quote date and the delivery date of a futures contract. We use \(M\) when referring to (discretized) monthly time periods (for example, a delivery month $M = $ Jan 2027). Importantly, a monthly delivery is a delivery \textit{at the end} of the month. For example, $F_t(M = \mathrm{Jan}\,2027)$ denotes the futures price at date $t$ for delivery at the end of January 2027.

\subsection{Designing early compute futures contracts}

Based on the current market and available data, the first working contracts will likely be designed around the following features:

\begin{enumerate}[leftmargin=*]
\item The underlying: a standardized GPU rental index.

The first contracts are expected to be launched on a current Silicon Data's \texttt{on-demand} rental index for the neo-cloud segment: H100, a central GPU benchmark, A100 or B200 (tickers \texttt{SDH100RT}, \texttt{SDA100RT}, and \texttt{SDB200RT}).

\item[2.] The settlement price and payoff.

The settlement price is the spot index value \(S_{T}\) at delivery time \(T\). The corresponding futures price is \(F_t(T)\). The settlement payoff of a long futures position entered at time $t$ at price \(F_t(T)\) is, therefore, 
\[
S_{T}-F_t(T).
\]

\item[3.] More practically, the first generation of compute futures may be settled against the \textit{average} daily spot index over the delivery month. We denote the monthly delivery contract by the superscript \(\mathrm{m}\). Let $d$ denote a specific day. The settlement price over month \(M\) is 
\[
S_M^{\mathrm{m}} = \frac{1}{N_M}\sum_{d\in M} S_d,
\]
with \(N_M\) representing the number of daily observations in month \(M\) and \(S_d\) representing a daily value of the spot index. The futures price quoted on day \(t\) for delivery in month \(M\) is denoted by \(F_t^{\mathrm{m}}(M)\). The corresponding settlement payoff is \(S_M^{\mathrm{m}}-F_t^{\mathrm{m}}(M)\).

\item[4.] As is natural, we are operating as if the underlying of a compute futures contract were 1 GPU-hour. That said, scaling up either the number of GPUs or the number of hours (or both) to determine contract values is straightforward.\footnote{For example, CME's cash-settled \href{https://www.cmegroup.com/content/dam/cmegroup/rulebook/NYMEX/13/1394.pdf}{PJM Western Hub peak monthly electricity futures contract} represents 5 MW (megawatts) during each of 16 peak hours per eligible weekday. Prices are quoted in dollars per MWh (megawatt-hour). For a month with 20 eligible peak days, the contract quantity is $5\times16\times20=1{,}600$ MWh. At an illustrative price of \$50 per MWh, its notional value is $50\times1{,}600=\$80{,}000$.} The proposed NYMEX rules in Submission 26-370, August 11, 2026, specify a 730 GPU-hour contract unit and business-day averaging for the H100 and B200 contracts. More broadly, guaranteeing transparency, reproducibility, representativeness and lack of an incentive for manipulation in the construction of a suitable index is delicate and involves trade-offs. Appendix \ref{comments}, which contains our filed response to the CFTC call for comments on the introduction of compute derivatives contracts, discusses these issues in the context of a simple model. For example, weighting individual providers' prices daily before daily and monthly index construction through averaging affects both manipulation incentives and representativeness. Similarly, monthly averaging may reduce manipulation sensitivity but it increases basis risk for users with daily exposures near expiry.

\end{enumerate}

Given this contract specification, the economic interpretation of the futures price would be the following: $F_t^{\mathrm{m}}(M)$ represents the agreed-upon price at date $t$ for average compute services over the delivery month $M$.

Monthly averaging is a reasonable way to structure the settlement mechanics given experience in the electricity market.\footnote{A PJM Western Hub peak calendar-month contract references electricity prices averaged over peak hours prices of the month rather than at one specific time instant.} Taking an average over the delivery interval is natural for a non-storable service flow like electricity or compute. A single point-in-time spot price may, in fact, contain excessive high-frequency fluctuations.

This said, specifying a generic point-in-time contract allows for cleaner continuous-time modeling while avoiding the complications of averaging and monthly discretization. The two versions of the contract are closely related and numerically similar. Our theoretical arguments, below, use the point-in-time version when needed for clarity.

Turning to P\&L and return accounting, according to standard futures convention, the marked-to-market cumulative P\&L for a long position entered on day \(t\) is
\begin{eqnarray}\label{ret}
F_{t'}^{\mathrm{m}}(M)-F_{t}^{\mathrm{m}}(M)
\end{eqnarray}
given any marked-to-market date \(t'\) from time \(t\) to the settlement month \(M\). The corresponding daily return assuming notional leverage is:
\[
    r_{t + 1} = \frac{F_{t+1}^{\mathrm{m}}(M)-F_{t}^{\mathrm{m}}(M)}{F_{t}^{\mathrm{m}}(M)}.
\]

We will use hold-to-maturity (essentially, standardized telescopic sums of Eq. (\ref{ret})) as well as constant-maturity returns in Section \ref{sec:empirical-analysis}. 

\section{The no-arbitrage pricing of compute futures}
\label{sec:no-arbitrage}

We turn to compute futures pricing. Our emphasis will be on the financial principles that are deployable for this new market.

\subsection{Storage-based no-arbitrage pricing}\label{storage}

We begin with a negative observation. In the case of compute futures, there is no natural no-arbitrage link between the \textit{current} spot rental price and the futures price. Because compute service flow is not storable, cash-and-carry no-arbitrage arguments cannot be conducted. A GPU-hour that is not used today cannot be carried forward and delivered, say, next month. As pointed out, compute is closer to electricity than to oil, metals, or other inventory assets. We will, instead, show that the relevant pricing links for compute futures derive from term rental contracts (Subsection~\ref{sec:term-rentals-synthetic-futures}) and from the forward-looking expectation of spot prices at expiration (Sections~\ref{sec:risk-pricing}).

We follow the typical convention of stating cash-and-carry no-arbitrage trades in terms of \textit{forward}, rather than \textit{futures}, prices. To this end, we will assume a frictionless futures market in which:
\begin{itemize}
\item[($i$)] there is no marking-to-market.  
\item[($ii$)] the futures contract is written on the same underlying as the forward contract (no basis risk). 
\end{itemize}
If Condition ($i$) and ($ii$) are satisfied, then futures and forward prices should coincide. If, in turn, forward prices are linked to \textit{current} spot prices, the same link would apply to futures prices. In Subsection \ref{sec:term-rentals-synthetic-futures}, we remove Condition ($ii$) (and preserve Condition ($i$) as a maintained assumption) to introduce a reasonable wedge between futures and forward prices. 

The standard no-arbitrage relation between the current spot price and the \textit{forward} price of a commodity assumes storability. For a storable commodity, the forward price is pinned down by the economic cost of buying the commodity today and carrying it to the delivery date. In continuous time:
\begin{eqnarray}\label{arb}
F_t(T)=S_t\exp[(r_t+c_t-y_t)(T-t)],
\end{eqnarray}
where \(S_t\) is the current spot price, \(r_t\) is the financing rate, \(c_t\) is the proportional storage-cost rate, and \(y_t\) is the \enquote{convenience yield}, i.e., the benefit of holding inventory. The rates are annualized and treated as constant over the holding period \([t,T]\).

Ignoring convenience benefits, let $K_t(T)=S_t\exp[(r_t+c_t)(T-t)]$ denote the total purchase-and-carry cost at delivery. If $F_t(T)>K_t(T)$, an arbitrageur can buy the commodity, finance its purchase and storage, and simultaneously sell it forward. At delivery, the resulting profit is $F_t(T)-K_t(T)$. Conversely, an inventory owner who does not need the commodity before delivery can sell it today, avoid storage expenses, and buy it back forward. If $F_t(T)<K_t(T)$, this strategy improves terminal wealth by $K_t(T)-F_t(T)$ relative to retaining the inventory.

The first strategy would tend to raise spot prices and lower forward prices until the arbitrage gain is eroded. The reverse strategy may, however, be unattractive to owners who value having the commodity available before delivery. The forward price may therefore remain below the purchase-and-carry benchmark, i.e., $F_t(T)<K_t(T)$. A positive convenience yield, $y_t>0$, in Eq. (\ref{arb}) captures this benefit of holding inventory.

As emphasized above, storage technology is precisely what compute rental service lacks. A GPU-hour is a flow of service at a point in time. If the capacity is not used today, today's GPU-hour is lost: compute cannot be stored in a warehouse, carried forward, and delivered next month. The current spot rental price \(S_t\) is therefore the price of immediate access to capacity, not the price of a durable object that can be physically moved through time. As a result, Eq. (\ref{arb}) cannot provide a pricing restriction for compute forwards and compute futures written on the same underlying. A useful no-arbitrage reference value may, instead, derive from contracts that already shift compute access across time, such as term rentals and reservation contracts, which we discuss next.

\subsection{No-arbitrage pricing against term rentals}
\label{sec:term-rentals-synthetic-futures}

The physical compute market has an existing mechanism for setting forward compute prices before a formal financial futures market is created: term rental contracts. Compute is often purchased through \texttt{Reserved} contracts that commit to capacity over a fixed term, from a couple of months to 3 years. The existing \textit{implied} forward market and the upcoming futures market should,  therefore, be linked by no-arbitrage relations.

Let \(\Pi_t(t\rightarrow T)\) denote the fixed physical term rental \enquote{rate}, quoted at \(t\), for compute access over the interval \([t,T]\). The rate is dollars per GPU-hour. It is fixed over an interval \([t,T]\). A futures contract, however, is a marginal financial claim at a future date $T$ or, more practically, a future month $M,$ in the case of monthly futures. Therefore, the non-arbitrage relationship between term rental rates and financial futures prices requires some translation (similar to the translation between zero-coupon bond rates and forward rates in fixed income markets).\footnote{Some arrangements are even closer to forward contracts than standard \texttt{Reserved} instances. For example, Amazon EC2 Capacity Blocks for ML allow users to reserve GPU-based accelerated computing instances for a future date and for a specified duration.} Below, we ignore time discounting. General expressions with time discounting are in Appendix~\ref{app:synthetic-forward-discounting}.

\begin{itemize}[leftmargin=*]
\item \textit{Synthetic futures prices are differences of adjacent term rental prices.} 
The difference between two term rental prices (i.e., time multiplied by the relevant rate) yields the \textit{synthetic} futures price for access over a future interval. Let \(M\) denote a delivery month identified by its settlement date. The \textit{synthetic} futures price for delivery month \(M\) is

\begin{equation}
F_t^{\mathrm{syn},\mathrm{m}}(M) =
(M - t) \Pi_t(t \rightarrow M)-
(M - t - 1)\Pi_t(t \rightarrow M-1).
\label{eq:synthetic-forward-monthly-nodiscount}
\end{equation}

In the continuous-time limit of the interval, the synthetic forward price can be defined as the marginal extension of the term rental price:
\begin{equation}
F_t^{\mathrm{syn}}(T)
=
\frac{\partial}{\partial T}
\left[
(T-t)\Pi_t(t\rightarrow T)
\right].
\label{eq:synthetic-forward-continuous-nodiscount}
\end{equation}

\item \textit{A strip of futures is equivalent to a financialized term rental rate.} 
As an alternative to signing a term rental contract, a user can replicate the same compute access by buying a strip of futures contracts for delivery at each date from \(t\) to \(T\). Consistent with Eq. (\ref{eq:synthetic-forward-continuous-nodiscount}), the market's implied term rental rate, i.e., $\Pi^{\mathrm{fin}}_t(t\rightarrow T)$, is the integral of the futures price strip:
\begin{equation}
\Pi^{\mathrm{fin}}_t(t\rightarrow T)
=
\frac{1}{T-t}
\int_t^T F_t(s)\,ds .
\label{eq:futures-implied-rental-strip}
\end{equation}
In the discrete (monthly) case, let \(t\) denote the quote date, \(m\) its month index, and \(M>m\) the terminal delivery month. The corresponding translation is:
\begin{equation}\label{eq:futures-implied-rental-strip-monthly}
\Pi^{\mathrm{fin},\mathrm{m}}_t(m\rightarrow M)=
\frac{1}{M-m}
\sum_{J=m+1}^{M} F_t^{\mathrm{m}}(J).
\end{equation}
\end{itemize}
In a world in which physical term rentals and financial futures are perfect substitutes for compute services, the following (equivalent) no-arbitrage restrictions would apply:
\begin{equation}\label{eq:noarb-synthetic-benchmark}
F_t(T)
=
F_t^{\mathrm{syn}}(T)
\quad
\textrm{and}
\quad
\quad
\Pi^{\mathrm{fin}}_t(t\rightarrow T)
=
\Pi_t(t\rightarrow T).
\end{equation}
The first restriction is in terms of the \textit{synthetic} futures price (essentially, the prevailing forward price) and the futures price. The second is in terms of the term rental rate and the futures-implied rate. 

Assume \(\Pi^{\mathrm{fin}}_t(t\rightarrow T)>\Pi_t(t\rightarrow T)\), i.e., the futures strip is expensive relative to the physical term rental. An arbitrageur can (1) buy the underpriced physical term rental capacity from the market, (2) sublet the acquired capacity spot over the rental interval and, (3) simultaneously, sell the expensive futures strip. For any \(s\in[t,T]\), the provider spends $F^{\textrm{syn}}_t(s)$, receives the spot rental revenue \(S_s\) from the physical rental market and the short futures payoff \(F_t(s)-S_s\) from the financial position. Assuming no basis risk, the spot rental price earned by the provider is the same index used to settle the futures contract. Thus, the two \(S_s\) terms cancel exactly and the combined arbitrage gain is \(\Pi^{\mathrm{fin}}_t(t\rightarrow T)-\Pi_t(t\rightarrow T)\). In a frictionless market, this arbitrage trade pushes down the futures strip - or raises the term rental rate - until the two ways of monetizing future capacity are equated.

Conversely, let \(\Pi^{\mathrm{fin}}_t(t\rightarrow T)<\Pi_t(t\rightarrow T)\), i.e., the physical term rental is expensive relative to the futures strip. The arbitrage trade runs in the opposite direction. An arbitrageur can (1) sell the overpriced physical term rental capacity, (2) buy the needed capacity spot over the rental interval and, (3) simultaneously, buy the cheap futures strip. For any \(s\in[t,T]\), the user receives $F^{\textrm{syn}}_t(s)$, pays the spot rental cost \(S_s\) in the physical rental market and receives the long futures payoff \(S_s - F_t(s)\) from the financial position. Assuming, again, no basis risk, the user is left with an arbitrage gain of $\Pi_t(t\rightarrow T)-\Pi^{\mathrm{fin}}_t(t\rightarrow T)$. In a frictionless market, this arbitrage trade pushes up the futures strip - or lowers the term rental rate - until they are equal.

The natural arbitrageurs are compute \textit{providers} who can easily intermediate between different compute markets. A similar arbitrage could, however, be conducted by anyone (not just compute providers), \textit{without} going through the spot market, if (1) the compute futures contract allowed physical delivery (like the physical contract) or, alternatively, (2) if the physical contract were cash-settled (like the futures contract). Condition (1) is not likely to be satisfied. Condition (2) is currently not satisfied. 

Even more importantly, there is \textit{basis risk}. Futures contracts transfer \textit{indexed} rental-price exposure while a physical term rental gives actual access to \textit{specific} compute services over the contracted term. We turn to this issue, i.e., basis risk, and the resulting limits to arbitrage. In other words, we remove Condition ($ii$) in Subsection \ref{storage} while preserving Condition ($i$).

\subsubsection{Limits to arbitrage}
\label{subsec:limits-to-arbitrage}

The frictionless no-arbitrage conditions in Eq. \eqref{eq:noarb-synthetic-benchmark} are not expected to hold exactly because physical term rentals and cash-settled financial futures are not perfect substitutes. The result is that the viability of the two arbitrages is asymmetric. 

The harder arbitrage direction is the second. If physical term rentals are expensive relative to the futures strip (i.e., \(\Pi^{\mathrm{fin}}_t(t\rightarrow T)<\Pi_t(t\rightarrow T)\)), an arbitrageur would ideally buy the capacity specified in the sold term rental agreement in order to meet the (sold) rental agreement. This capacity, however, is with a particular provider, in a particular location, with a particular configuration, reliability tier, service relationship, and so on. The futures contract is, instead, written on a compute index. Thus, the spot payment to source capacity and the spot payment in the futures payoff do not cancel cleanly, in general. The arbitrageur remains exposed to basis risk. 

A similar concern exists regarding the first arbitrage strategy. The likely arbitrageur (a provider, as said) is, however, better positioned to mitigate it. The provider now buys under the term rental agreement before selling spot. The provider controls capacity across customers, regions, configurations, and service tiers. It can drop the acquired capacity in a pool and effectively perform some degree of indexation internally. For this reason, the spot inflow from sold capacity can be close to the spot outflow under the futures contract. The corresponding basis risk is, therefore, expected to be lower than in the previous arbitrage. 

We conclude that the no-arbitrage relationship between $\Pi_t(t\rightarrow T)$ and $\Pi^{\mathrm{fin}}_t(t\rightarrow T)$ is likely to hold with a \enquote{physical access wedge}:
\begin{align}
\Delta^\Pi_t(t\rightarrow T) = 
\Pi_t(t\rightarrow T)
-
\Pi^{\mathrm{fin}}_t(t\rightarrow T).
\label{eq:physical-access-premium}
\end{align}
Applying the translation from term rental rates to synthetic futures, we also have:
\begin{align}\label{eq:physical-access-premium-forward}
F_t(T)
=
F_t^{\mathrm{syn}}(T)
-
\Delta^F_t(T),
\end{align} 
where
\begin{align}
\Delta^F_t(T)
=
\frac{\partial}{\partial T}
\left[
(T-t)\Delta^\Pi_t(t\rightarrow T)
\right].
\end{align}

We expect the physical access wedge to be:
\begin{itemize} 
\item non-negative, i.e., $\Delta^\Pi_t(t\rightarrow T) \geq 0$, due to the described asymmetry in the success of the two arbitrage strategies.  
\item increasing in the time horizon \(T-t\). Over a short period, the physical access wedge is driven by idiosyncratic providers' characteristics as well as localized friction, capacity bottlenecks, SLA delivery mismatches and so on. In essence, the spot price for a specific server box may decouple from a generalized network index. Over a long period, the arbitrageur is not just dealing with providers' idiosyncrasies. She/he is also dealing with technological change. The physical hardware sold under the rental agreement today becomes progressively obsolete, while the financial index is likely to, e.g., dynamically rebalance its underlying basket to track next-generation chips. 
\end{itemize}

Dispensing with institutional issues associated with marking-to-market in the futures market, we conclude that \textit{synthetic} futures (i.e., implied forward) prices are likely to be upper bounds on the prices of financial futures:
\begin{equation}
F_t(T)
<
F_t^{\mathrm{syn}}(T).
\label{eq:noarb-synthetic-upper-bound}
\end{equation}

We may, however, expect the tracking gap to decrease over time, as compute intermediaries and providers offer more standardized capacity packages that are closer to the index specification.

\section{The risk pricing of compute futures}
\label{sec:risk-pricing}

We may draw lessons about reasonable compute risk pricing from the electricity forward markets. \citet{bessembinder2002equilibrium} and \citet{longstaff2004electricity} document how the electricity forward premium (i.e., the difference between expected spot value and forward values, i.e., $ \lambda^{\mathrm{syn}}_t(T)=\mathbb{E}_t[S_T - F^{\mathrm{syn}}_t(T)]$) can switch sign depending on the side of the market (providers relative to users) exerting the stronger hedging pressure. Electricity providers are exposed to revenue risk from spot variance, while users are especially exposed to positive price spikes (right skewness). The former has a \textit{positive} effect on the premium since providers would sell forward (thereby reducing $F^{\mathrm{syn}}_t(T)$ in equilibrium) in order to benefit from downward price moves. The latter has a \textit{negative} effect on the premium since users would buy forward (thereby increasing $F^{\mathrm{syn}}_t(T)$ in equilibrium) in order to benefit from upward price moves. Analogous dynamics are likely at play for compute. 

Next, we discuss broader implications from the upcoming financialization of compute. Beyond allowing compute market participants, like cloud providers and AI developers, to hedge operational risks, the financialization of compute would, in principle, allow financial market participants to take bets on the future of the AI economy. The result should be superior compute price discovery and, in the end, improved efficiency in compute investment and financing decisions.

The asset-pricing framework used to formalize the risk-pricing of traditional financial assets, such as stocks, bonds and derivatives, should apply. The overarching principle is that asset prices are \enquote{discounted} expectations of future payoffs in which the discounting takes time and risk into account. We have
\[
F_t(T)
= \mathbb{E}_t[S_T]
- \lambda_t(T),
\]
where $\mathbb{E}_t[S_T]$ is the expected future spot rental price and $\lambda_t(T)$ is the risk premium for holding a long position in compute futures contracts. The risk premium can, therefore, be expressed as 
\[
    \lambda_t(T) = \mathbb{E}_t[S_T - F_t(T)].
\] 
The quantity $\lambda_t(T)$ is a ``risk premium'' in the sense that it is the expected payoff of a zero-upfront-cost long future position, i.e., it is the compensation that investors require for buying compute forward.\footnote{The risk-neutral or $\mathbb{Q}$-expectation of the future payoffs $S_T - F_t(T)$ would, on the other hand, be zero, i.e.,
\[ 
0 = \mathbb{E}^\mathbb{Q}_t[S_T - F_t(T)],
\]
which leads to 
\[
F_t(T) = \mathbb{E}^\mathbb{Q}_t[S_T].
\]
Compute futures prices are $\mathbb{Q}$-expectations of compute spot prices at delivery.}

Canonically, the risk premium is also the (standardized) covariance between the compute payoff and the stochastic discount factor, i.e. $M_{t,T}$:\footnote{Appendix~\ref{app:risk-premium-sdf} provides the detailed derivation.}
\[
\lambda_t(T)
=
-\frac{\operatorname{Cov}_t(M_{t,T},S_T)}{\mathbb{E}_t[M_{t,T}]},
\]
with $M_{t,T}$ an increasing function of the marginal utility of the \enquote{marginal} compute futures investor.

Assume compute providers are the primary hedgers and the marginal investor. The covariance between compute prices and their marginal utility is expected to be negative since providers benefit from higher prices.\footnote{Assume the terminal wealth of the compute providers, $W_T,$ is given by $Q_TS_T,$ where $Q_T$ is the number of GPU-hours sold. Then, their marginal utility of wealth would be equal to $u^{\prime}(W_T) = u^{\prime}(Q_TS_T).$ Under risk aversion, the marginal utility would decrease with increases in $S_T.$} The producers are short hedgers and the long positions taking the opposite side of their trades provide \enquote{insurance} to them. In equilibrium, compute risk premia are positive because the long positions ought to be compensated for the insurance they provide. Should the primary hedgers be AI developers, the logic would be the same but, of course, reversed.

Upon the initial launch, we expect compute futures market to be dominated by natural hedgers (i.e., compute-service providers and AI developers). Given our evidence on a \textit{positive} compute risk premium provided in Section \ref{sec:empirical-analysis}, we expect the short selling decisions of compute providers to be particularly impactful at the onset.

In a mature futures market, however, hedge funds, commodity trading advisors, commodity index funds, proprietary trading firms and asset managers, among others, would play a role in leading to equilibrium prices. The sign of the risk premium would, again, depend on whether the price of compute is high in \enquote{good} or in \enquote{bad} states of nature for the marginal investor. Based on the current growth in AI adoption, let us assume that good states for financial markets are those in which a growing AI infrastructure is broadly viewed as boosting profitability. In this scenario, compute prices are likely to be relatively higher and the correlation between $M_{t,T}$ and $S_T$ would be negative. The risk premium would, therefore, be positive ($\lambda_t(T) > 0$). Intuitively, because compute prices are high in good states and low in bad states, a long future position is risky (it pays when a payment is not needed) and, therefore, should be priced at a \textit{discount}: 
\begin{eqnarray}\label{rel1}
F_t(T) < \mathbb{E}_t[S_T].
\end{eqnarray} 
A positive sign for the compute risk premium hinges on a demand-side argument. The development of the AI economy will, also, be driven by supply-side innovations, such as breakthroughs in semiconductor technology lowering the cost of compute and rendering existing GPU benchmarks obsolete. In this scenario, the sign of the compute risk premium may even be negative. 

We have presented qualitative arguments regarding the potential sign of the compute risk premium. We have done so: 
\begin{itemize}
\item In an economy in which the main hedging motive is operational, i.e., the current economy. 
\item In an economy in which mature financialization of compute may lead to a broader hedging role for compute akin to that of traditional asset classes, i.e., a prospective economy in which compute is an established asset class.
\end{itemize}
In the latter case, should the marginal investor be an equity market investor, the pricing factors that have been put forward for valuation in the equity market (i.e., the Fama-French factors and many others) would be natural candidates for understanding risk compensation. Not only are we at the infancy of compute financialization as a project let alone a reality, time will also be needed for some degree of integration between equity and compute markets to justify \textit{joint} pricing.

We are - and will continue to be in the short-to-medium term after the launch of compute futures - in the first economy, one in which the main hedging motive is operational. In light of these observations, it would be economically inappropriate - and hardly informative given available data -  to deploy the typical asset-pricing arsenal (through, i.e., the computation of forward-looking covariances given assumed pricing factors) to compute data. We will, instead, use spot and \textit{synthetic} futures prices to provide empirical evidence on the current sign and magnitude of the term structure of \textit{synthetic} compute risk premia \textit{without} assuming an ad-hoc asset-pricing model, i.e., by solely computing average hold-to-maturity and fixed-maturity (excess) returns (c.f. Section \ref{sec:empirical-analysis}).  

We conclude this section by emphasizing that we were careful in emphasizing that a positive compute risk premium is not a statement about positive correlation between compute prices and the state of the economy as a whole. In the current economy, it is a statement about more aggressive hedging on the part of compute-service providers relative to users. After mature financialization, it would be a statement about the impact of compute on the utility of the marginal investor. Whether such a utility positively correlates with the state of the broader economy should be the subject of discussion regarding the societal impact of widespread AI adoption. 

\section{Empirical analysis}
\label{sec:empirical-analysis}

We now turn to an empirical analysis of the risk-return properties of prospective compute futures. We employ spot prices and term rental prices provided by Silicon Data. Similarly to the spot prices described in Subsection \ref{index1}, the term rental prices are term rental \textit{indexes} and, in essence, averages across idiosyncratic capacities. The resulting physical access wedge is, also, an average across idiosyncratic physical access wedges. We write:
\begin{eqnarray*}
\overline{\Delta}^\Pi_t(t\rightarrow T) = \overline{\Pi}_t(t\rightarrow T) - \Pi^{\mathrm{fin}}_t(t\rightarrow T) > 0,
\end{eqnarray*}
and, equivalently,
\begin{eqnarray*}
\overline{F}_t^{\mathrm{syn}}(T) = F_t(T) + \overline{\Delta}^F_t(T),
\end{eqnarray*}
with $\overline{\Delta}^F_t(T)>0.$

We use \textit{synthetic} futures prices (i.e., indexes), $\overline{F}_t^{\mathrm{syn}}(T)$, as proxies for $F_t(T).$ The measurement error is additive, positive and equal to the physical access wedge $\overline{\Delta}^F_t(T).$ The approach results in a ``rehearsal'' empirical environment in which to study the risk-return properties of prospective compute futures before actual contracts are launched and traded.
  
We focus on the \textit{sign} and \textit{magnitude} of the estimated \textit{compute risk premium}. Our early evidence points to a (largely) positive risk premium which we interpret as the outcome of relatively more aggressive hedging on the part of compute-service providers. Refinements will hinge on market developments and, of course, on the future availability of exchange-traded futures data.
  
We begin by describing the term rental and \textit{synthetic} futures data (Subsection~\ref{sec:silicon-forward-data}). We are able to construct a return panel in two dimensions: compute generation (A100, H100, B200) and maturity. We use the data to define hold-to-maturity returns (Subsection \ref{sec:hold-to-maturity-payoff}) and rolling constant-maturity returns (Subsection \ref{sec:constant-maturity-return}).

\subsection{Term rentals and synthetic futures prices}
\label{sec:silicon-forward-data}

Silicon Data compiles term structure curves from observed term rental agreements.\footnote{Details regarding data handling are discussed in Appendix \ref{data1} and in Appendix \ref{data2}.} The data covers three GPU benchmarks (A100, H100, B200). We have tenors from zero to 36 months that are interpolated on a 0.25-month grid.\footnote{We replicated Silicon Data's synthetic forward curve construction using a continuous term structure without discounting, c.f. Eq. \eqref{eq:synthetic-forward-continuous-nodiscount}, with only minor discrepancies.} The data frequency is daily and matches the spot index.  

\begin{figure}[!t]
\centering
\includegraphics[width=0.92\linewidth]{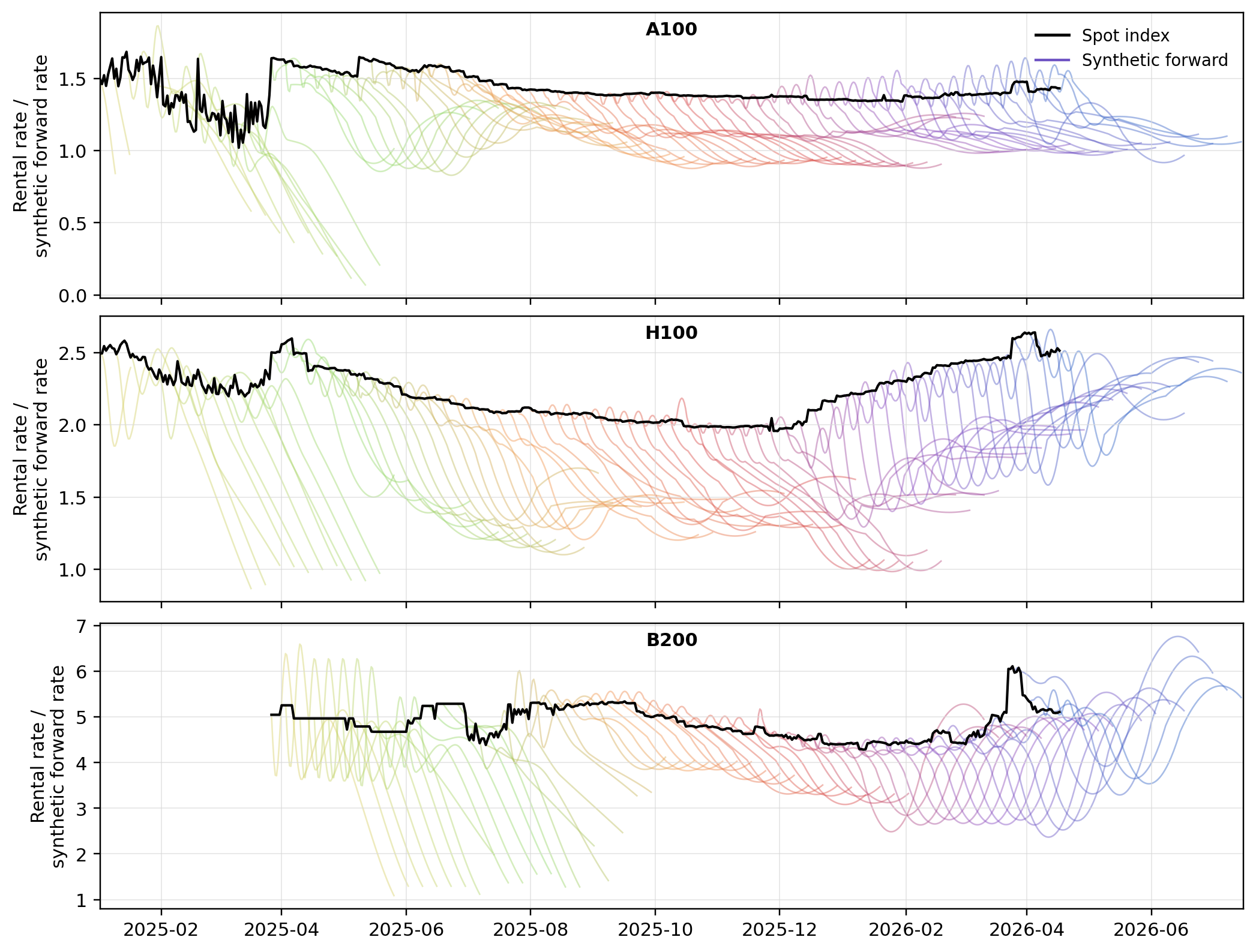}
\caption{\textbf{Silicon Data forward curves for GPU rental benchmarks.} The black line is the zero-tenor value of Silicon Data's forward curve, which closely tracks the spot rental index in the later part of the sample period (c.f., Fig. \ref{fig:index}). The colored curves are synthetic forward curves from Silicon Data sampled at seven-day frequency. Each forward curve has a three-year tenor. The curves are horizontally compressed in order to display the full three-year shape.}\label{fig:silicon-forward-curves}
\end{figure}

Fig.~\ref{fig:silicon-forward-curves} plots the forward curves for the three GPU benchmarks sampled using a seven-day frequency. The short end of the forward curves tracks closely the spot index (as also shown in Fig. \ref{fig:index}).\footnote{In Appendix~\ref{app:zero-tenor-forward}, we show that the zero-tenor rental strip coincides with the spot (rental) rate.}
The forward curves are downward sloping over most of the time period. Their shape reflects the mild downward trend in the spot prices and the expectation that rental prices will decline with technology innovation, new supply and obsolescence of existing GPU generations. Starting in late 2025, however, the forward curves begin to curve upward at longer horizons, especially for the more advanced H100 and B200 benchmarks. This pattern is especially strong in the case of B200, for which the long end of the forward curve is particularly suggestive of the expectation of a sharp increase in future rental prices reflecting increased demand for frontier compute. This phenomenon is less pronounced for A100, likely because older GPU generations are less impacted by the most recent wave of AI demand.

\subsection{Hold-to-maturity returns and risk premium}
\label{sec:hold-to-maturity-payoff}

We first study the most natural objects: hold-to-maturity returns. For a long position entered at date \(t\) and held to delivery month \(M\), the payoff is
\[
S_M-\overline{F}^{\mathrm{syn}}_t(M),
\]
where \(S_M\) is the spot rental index for delivery month \(M\) and, once more, \(\overline{F}^{\mathrm{syn}}_t(M)\) is the \textit{synthetic} futures price inferred from the rental term structure at date \(t\). We report the hold-to-maturity return as the payoff normalized by the notional value of the contract: 
\[
r_{t\to M}=
\frac{S_M-\overline{F}^{\mathrm{syn}}_t(M)}
{\overline{F}^{\mathrm{syn}}_t(M)}.
\]
The working assumption is that entering the synthetic futures position at date \(t\) requires upfront capital \(F_t(M)\). In practice, marking-to-market requires less upfront capital. Thus, the actual return on posted capital would be a leveraged version of this return. 

Given this construction,  the realized return can be decomposed as follows
\[
r_{t\to M}
=
\frac{\overline{\lambda}^{\mathrm{syn}}_t(M)}{\overline{F}^{\mathrm{syn}}_t(M)}
+
\varepsilon_{t\to M},
\qquad
\text{ with    }
\qquad
\varepsilon_{t\to M}
:=
\frac{S_M-\mathbb{E}_t[S_M]}
{\overline{F}^{\mathrm{syn}}_t(M)}.
\]
The first term is the risk premium $\overline{\lambda}^{\mathrm{syn}}_t(M)$ normalized by the futures notional. The second term is a martingale difference sequence shock (i.e., a shock with zero conditional mean, $\mathbb{E}_t[\varepsilon_{t\to M}]=0$). We estimate the risk premium by simply averaging the realized returns across the sample.

\begin{figure}[!t]
\centering
\vspace{-0.28cm}
\includegraphics[width=\linewidth,trim=20 12 42 42,clip]{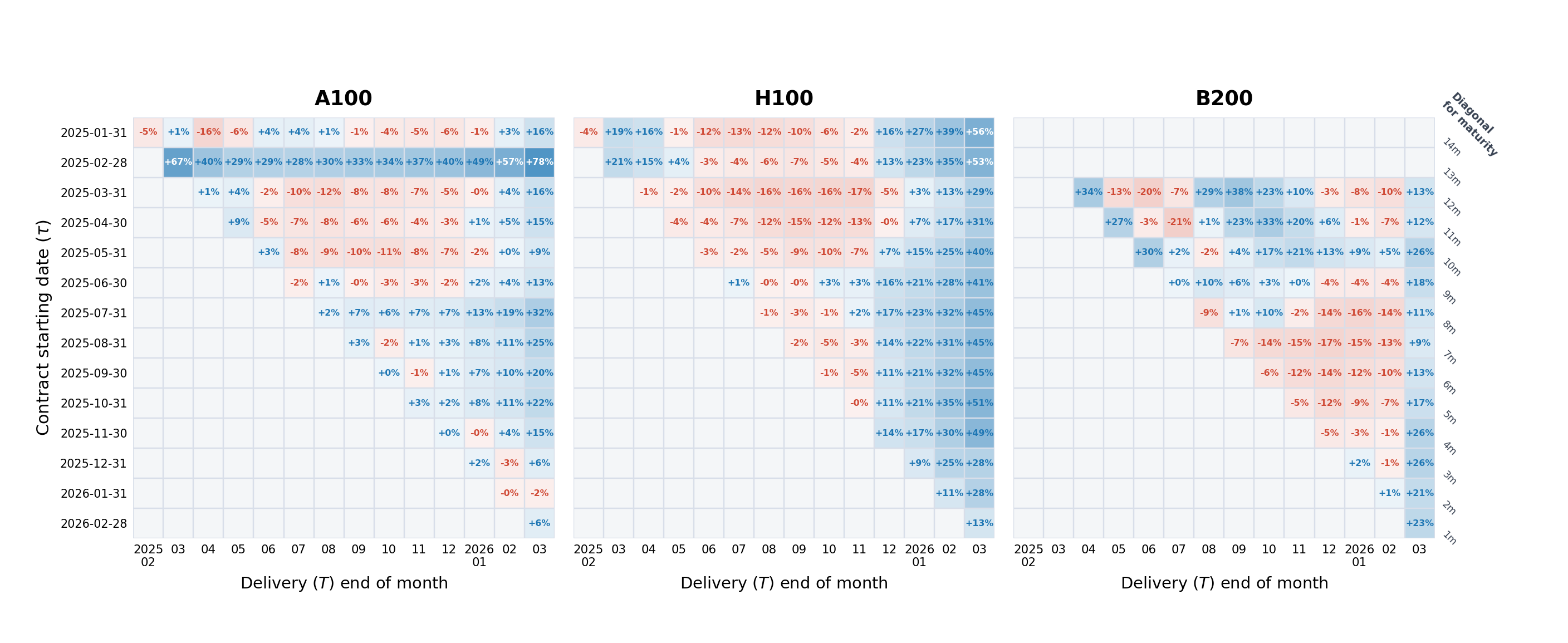}
\vspace{-0.15cm}
\caption{\textbf{Hold-to-maturity returns of synthetic compute futures.} The figure reports realized long hold-to-maturity returns (payoff normalized by notional value), i.e., \(r_{t\to M}=(S_M-\overline{F}^{\mathrm{syn}}_t(M))/\overline{F}^{\mathrm{syn}}_t(M)\). Rows are contract starting dates \(t\) and columns are delivery months \(M\). \(S_M\) is measured using end-of-month spot observations.}
\label{fig:hold-to-maturity-payoff}
\end{figure}

Fig.~\ref{fig:hold-to-maturity-payoff} reports the hold-to-maturity return matrix. Rows are contract starting dates and columns are delivery months. The triangular structure comes from the obvious requirement that delivery must occur after the contract starting date. Each diagonal contains contracts with the same maturity (or returns with the same holding period length).

The figure shows both positive and negative returns. The positive returns, however, are more prevalent and tend to be larger in magnitude, especially for later maturity months in the sample.

Turning to the risk premium estimates, Table~\ref{tab:hold-to-maturity-payoff-means} reports selected maturity/diagonal averages after dropping the first two start-date rows in Fig.~\ref{fig:hold-to-maturity-payoff}. As mentioned, the early records are the least reliable. Average returns (and estimated premia) are positive for all three GPU generations. In raw units, the equally-weighted averages across maturities 1 through 12 months are \(4.2\%\) for A100, \(13.4\%\) for H100 and \(4.3\%\) for B200. In annualized terms, the corresponding averages are \(7.5\%\), \(26.2\%\) and \(11.2\%\). Notwithstanding rare exceptions, the maturity breakdown is also consistent with positive compute risk premia. The longer maturities tend to have larger average returns, although the pattern is not monotonic. Because of their reliance on a small number of overlapping observations, the long-maturity estimates are especially not robust.

\begin{table}[!t]
\centering
\footnotesize
\setlength{\tabcolsep}{2pt}
\begin{tabular}{@{}>{\raggedright\arraybackslash}p{0.20\linewidth}
*{3}{>{\centering\arraybackslash}p{0.085\linewidth}}
@{\hspace{0.08\linewidth}}
*{3}{>{\centering\arraybackslash}p{0.085\linewidth}}@{}}
\toprule
& \multicolumn{3}{c}{Panel A: average returns (\%)}
& \multicolumn{3}{c}{Panel B: annualized  (\%)} \\
\cmidrule(lr){2-4}\cmidrule(lr){5-7}
maturity $\times$ generation & A100 & H100 & B200 & A100 & H100 & B200 \\
\midrule
3m  & 0.83  & 6.54  & -3.83 & 3.32  & 26.16 & -15.31 \\
6m  & 2.97  & 11.40 & 10.21 & 5.94  & 22.80 & 20.41  \\
9m  & 2.18  & 17.17 & 4.58  & 2.91  & 22.89 & 6.11   \\
12m & 16.34 & 28.72 & 12.61 & 16.34 & 28.72 & 12.61  \\[6pt]
All-in average & 4.18 & 13.43 & 4.28 & 7.46 & 26.22 & 11.16 \\
\bottomrule
\end{tabular}
\vspace{0.15cm}
\caption{\textbf{Average hold-to-maturity returns.} The table reports average realized long hold-to-maturity returns, i.e., \(r_{t\to M}=(S_M-\overline{F}^{\mathrm{syn}}_t(M))/\overline{F}^{\mathrm{syn}}_t(M)\). We take the average across the diagonal in Fig.~\ref{fig:hold-to-maturity-payoff}. We start the sample from 2025-03 by dropping the first two start-date rows in Fig.~\ref{fig:hold-to-maturity-payoff} to accommodate questionable data quality. Panel A reports average returns. Panel B annualizes each maturity by multiplying by \(12/h\). The ``All-in average'' row is the equally-weighted average across maturities \(h=1,\ldots,12\).}
\label{tab:hold-to-maturity-payoff-means}
\end{table} 

Although our numbers should only be interpreted as first-pass estimates from a necessarily short sample over a time horizon with considerable AI growth, we document positive compute risk premia in the high single to double digits. As a rule of thumb for pricing futures, if the expected B200 spot rental price one year ahead is \(\$7\) dollars per GPU-hour, a futures price around \(\$6.25\) would not be an unreasonable risk-adjusted price. It would correspond to a positive risk premium of about \((\$0.75 \div \$6.25) = 12\%\) of the futures notional, approximately equal to the 12.61\% estimate in Table~\ref{tab:hold-to-maturity-payoff-means}.

The hold-to-maturity analysis is useful to assess reasonable futures values ($F_t(T)$) directly from expected compute prices at the delivery date ($\mathbb{E}_t[S_M]$). Equivalently, it is helpful to quantify the term structure of compute risk. However, investors may not always intend to hold every contract to maturity and deal with maturity drifts towards zero. For this reason, we next construct rolling constant-maturity returns and study the corresponding risk-return properties. 

Additionally, we note that the hold-to-maturity analysis above stops at the longest completed maturity of 12 months (or 14 months for A100 and H100), which is the longest life of a contract in our sample. However, our data contains forward curve information up to a three-year maturity which is ignored in the above hold-to-maturity analysis. The constant-maturity return approach in Subsection \ref{sec:constant-maturity-return} below allows us to study the risk-return properties of the longer-maturity contracts as well by continuously rolling the position in order to maintain a constant maturity of, e.g., 24 or 36 months.

\subsection{Constant-maturity daily returns and risk premium}
\label{sec:constant-maturity-return}

We construct a return panel of continuously-rolled \textit{daily} constant-maturity long \textit{synthetic} futures returns. The strategy rolls the delivery month forward as time passes in order to set a target maturity (e.g., 24 months). We assume that the rolling is done at the end of each month, meaning that the maturity is a fixed target up to a drift of at most one month. 

The construction has two ingredients: 1) the daily return series for a \textit{synthetic} futures contract \(r_{t}^{[M]}\) with fixed delivery month \(M\) (e.g., $M = ~$Jan 2029) and 2) a constant-maturity return series \(r_{t}^{(h^{\mathrm{m}})}\) that rolls the delivery month $M$ forward to maintain a constant maturity of \(h^{\mathrm{m}}\) (e.g., $h^{\mathrm{m}} = 24$ months). Specifically:
\begin{enumerate}
\item[1)] For a fixed delivery month \(M\), the daily \textit{synthetic} return on trading day \(t\) is
\[
r_{t}^{[M]}
:=
\overline{F}^{\mathrm{syn}}_{t}(M) /
\overline{F}^{\mathrm{syn}}_{t-1}(M)
-1.\footnotemark
\]
\footnotetext{Here, \(t\) indexes trading days and \(t-1\) denotes the previous trading day. In the empirical implementation, trading days are defined using the \texttt{CME\_TradeDate} calendar from \texttt{pandas-market-calendars}. The return is measured from the end-of-day futures curve on trading day \(t-1\) to the end-of-day futures curve on trading day \(t\).}%
\item[2)] Fixing a constant maturity \(h^{\mathrm{m}}\), the strategy targets the delivery month that is \(h^{\mathrm{m}}\) months ahead of the current month, that is: \(M = \operatorname{month}(t)+h^{\mathrm{m}}\), where \(\operatorname{month}(t)\) is the month of trading day \(t\). The constant-maturity return is then:
\[
r_{t}^{(h^{\mathrm{m}})}
:=
r_{t}^{\left[\operatorname{month}(t)+h^{\mathrm{m}}\right]}
=
\overline{F}^{\mathrm{syn}}_{t}
\!\left(
\operatorname{month}(t)+h^{\mathrm{m}}
\right) /
\overline{F}^{\mathrm{syn}}_{t-1}
\!\left(
\operatorname{month}(t)+h^{\mathrm{m}}
\right)
-1.
\]
Within a month, this is the one-day marked-to-market return on the same delivery-month contract. At the beginning of a new month, the strategy rolls into the new \(h^{\mathrm{m}}\)-month-ahead delivery month, and the return is measured against the previous trading day's quote for that newly targeted contract.
\end{enumerate}

The constant-maturity return series \(r_{t}^{(h^{\mathrm{m}})}\) could, in principle, be used to estimate the \textit{per-period} compute premium for a position with a given fixed maturity. The hold-to-maturity risk premium can, in fact, be decomposed into the sum of \textit{per-period} risk premia. 

In order to make the decomposition additive, we work with log returns denoted by tildes. For a contract entered in month \(m\) for delivery month \(M\), the hold-to-maturity log return
\(
\widetilde{r}_m(M)
:=
\log S_M-\log \overline{F}^{\mathrm{syn}}_m(M)
\)
can be decomposed into the sum of monthly log returns of decreasing maturity:
\[
\widetilde{r}_m(M)
=
\widetilde{r}_{m+1}^{(M-m-1)}
+\widetilde{r}_{m+2}^{(M-m-2)}
+\cdots
+\widetilde{r}_{M}^{(0)},
\]
where
\(
\widetilde{r}_{m}^{(h)}
:=
\log F^{\mathrm{syn}}_{m}(m + h)-\log F^{\mathrm{syn}}_{m-1}(m + h)
\)
is the log return accumulated during month \(m\) with residual maturity of length \(h\). Taking conditional expectations at date \(m\), the log-return compute premium has the following decomposition:
\begin{align*}
\widetilde{\lambda}^{\mathrm{syn}}_m(M)
:=
\mathbb{E}_m[\widetilde{r}_m(M)]
=
\mathbb{E}_m
\left[
\widetilde{\mu}_{m}^{(M-m-1)}
+\widetilde{\mu}_{m+1}^{(M-m-2)}
+\cdots
+\widetilde{\mu}_{M-1}^{(0)}
\right],
\end{align*}
where \(\widetilde{\mu}_{s-1}^{(h)}:=\mathbb{E}_{s-1}[\widetilde{r}_{s}^{(h)}]\) is the per-period log-return premium for a return accumulated during month \(s\). Assuming that these per-period premia are constant over time ($m$), we have
\begin{equation}
\widetilde{\lambda}^{\mathrm{syn}}_m(M)
=
\widetilde{\mu}^{(M-m-1)}
+\widetilde{\mu}^{(M-m-2)}
+\cdots
+\widetilde{\mu}^{(0)}.
\label{eq:log-return-premium-decomposition}
\end{equation}
Each \(\widetilde{\mu}^{(h)}\) can, therefore, be estimated by a time-series average of log returns for the corresponding constant-maturity strategy:
\[
\widehat{\widetilde{\mu}}^{(h)}
=
\frac{1}{|\text{available sample}|}
\sum_{t\in\text{available sample}}
\widetilde{r}_{t}^{(h)}.
\]

\begin{figure}[!t]
\centering
\includegraphics[width=\linewidth]{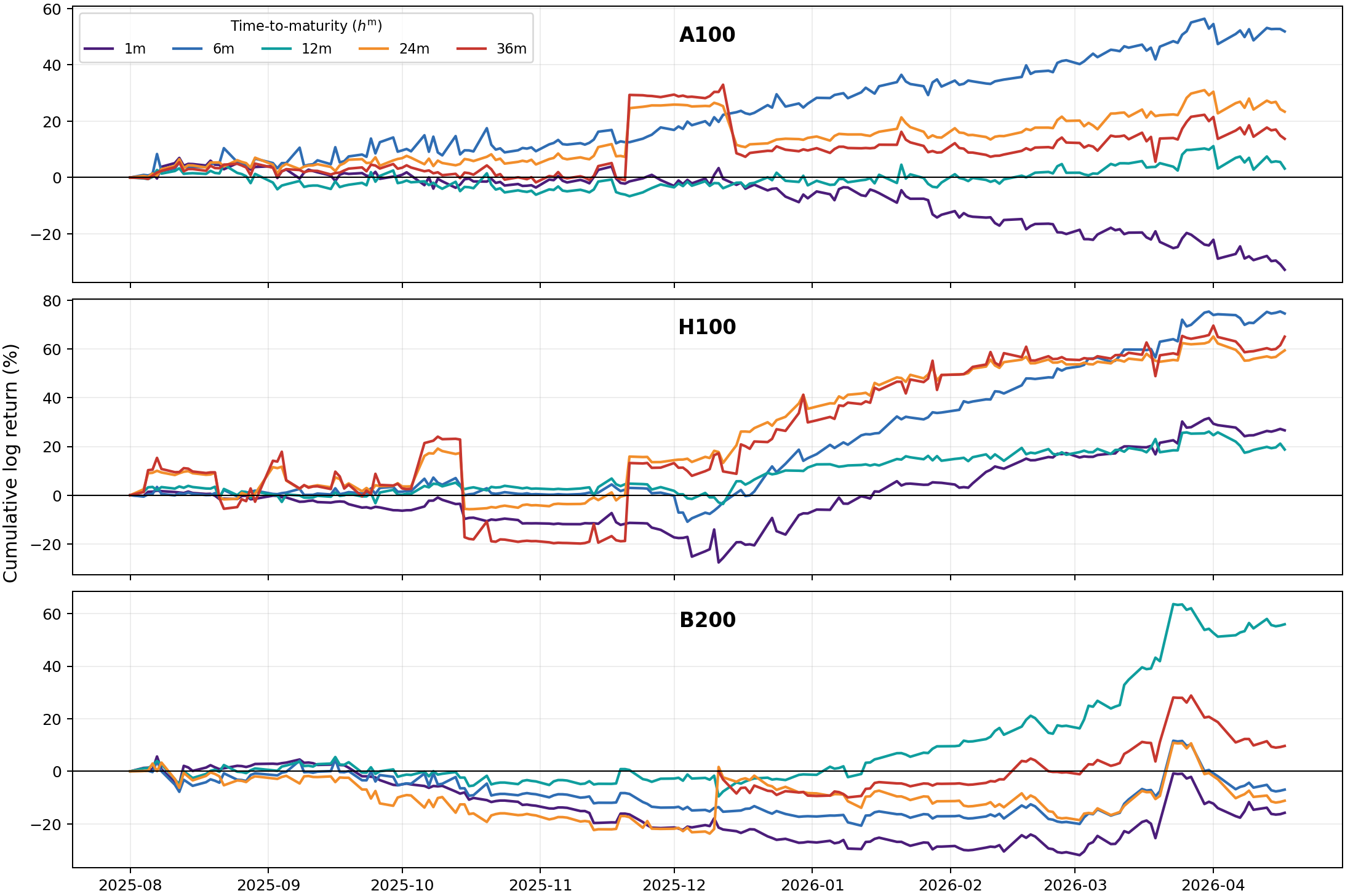}
\caption{\textbf{Cumulative returns on rolling constant-maturity positions.} Cumulative daily log returns on rolling constant-maturity synthetic forward positions: $\sum_{t} \log(1+r_{t}^{(h^{\mathrm{m}})})$. The sample period begins on August 1, 2025, except for the B200 36-month strategy which starts on December 11, 2025.}
\label{fig:constant-maturity-cumulative}
\end{figure}

Fig.~\ref{fig:constant-maturity-cumulative} plots the cumulative log returns of the rolling constant-maturity strategy for the three GPU generations and for \(h^{\mathrm{m}}=1, 6, 12, 24, 36\) months. Table~\ref{tab:risk-return-double-sort} reports mean returns, standard deviations and market betas for the same quantities.

The sample's start date is August 1, 2025 for each return series, which is the date when the B200 forward curve becomes robust. For A100 and H100, the return series appears to be reliable at a somewhat earlier date starting around April 1, 2025, when the zero-tenor value of the forward curve becomes consistent with the spot index. This said, we align the sample start date across the three GPU benchmarks to produce a cleaner comparison across the three generations. The only exception is the B200 36-month strategy, which starts on December 11, 2025.\footnote{In Appendix~\ref{app:full-sample-returns}, we report the return series over the longest available sample, inclusive of time periods with poor data quality.} 

The cumulative return paths in Fig.~\ref{fig:constant-maturity-cumulative} behave reasonably, with short-run fluctuations and persistent trends. In this sense, they resemble familiar charts in technical analysis. The paths show both co-movements and visible spreads across maturities.

Table~\ref{tab:risk-return-double-sort} reports risk and return summary statistics for the constant-maturity strategies. The mean and the standard deviation are annualized and estimated from the same sample in Fig.~\ref{fig:constant-maturity-cumulative}. The market betas are estimated against the daily Fama-French market excess return. 

Mean returns are more often positive than negative: 11 of the 15 compute-generation/maturity strategies have positive annualized mean returns. Long positive compute \textit{synthetic} futures appear, on average, to earn a positive compensation for bearing compute price risk. The main exceptions are at the short end of the curve. The one-month strategy is negative for A100 and B200 and is the least profitable maturity for both generations. For H100, the one-month strategy is positive. It is the second least profitable maturity in this case. 

\begin{table}[!t]
\centering
\small
\begin{tabular}{lrrr@{\hspace{0.75cm}}rrr@{\hspace{0.75cm}}rrr}
\toprule
\multicolumn{1}{c}{}
& \multicolumn{3}{c}{mean return (\% ann.)}
& \multicolumn{3}{c}{std. return (\% ann.)}
& \multicolumn{3}{c}{market beta} \\
\cmidrule(lr){2-4}
\cmidrule(lr){5-7}
\cmidrule(lr){8-10}
maturity $\times$ generation
& A100 & H100 & B200
& A100 & H100 & B200
& A100 & H100 & B200 \\
\midrule
1m  & -41.7 & 44.4 & -15.3 & 30.2 & 36.5 & 38.5 & 0.21 & -0.11 & 0.18 \\
6m  & 82.3 & 111.6 & -1.7 & 42.0 & 33.9 & 41.3 & 0.34 & 0.05 & 0.43 \\
12m & 9.3 & 30.7 & 88.2 & 31.1 & 29.0 & 43.5 & 0.48 & 0.25 & 0.29 \\
24m & 40.0 & 97.2 & -2.3 & 37.6 & 50.5 & 54.2 & 0.29 & 0.12 & 0.11 \\
36m & 34.9 & 131.7 & 39.4 & 57.7 & 89.2 & 48.4 & 0.43 & 0.12 & 0.33 \\
\bottomrule
\end{tabular}
\vspace{0.15cm}
\caption{\textbf{Risk and return across compute generations and maturity.} The table reports rolling constant-maturity returns sorted by time-to-maturity (rows) and GPU generation (columns). Mean returns are annualized by multiplying the daily mean returns by 252. Standard deviations are annualized by multiplying the daily standard deviations by \(\sqrt{252}\). Both are reported in percent. The market betas are the OLS slope coefficients from regressing the daily compute forward excess returns on the daily Fama-French market excess returns. The current Fama-French daily file is updated only through March 31, 2026, while the Silicon return sample runs through April 17, 2026. The beta estimates therefore miss 12 trading days.}\label{tab:risk-return-double-sort}
\end{table}

Importantly, the short-maturity underperformance may not reflect the true compute futures risk premium but a ``run-off'' of the physical access wedge embedded in the \textit{synthetic} futures. The physical access wedge is, in fact, expected to converge to zero as maturity approaches zero: $\lim_{T \downarrow t} \overline{\Delta}^\Pi_t(t\rightarrow T) = 0.$ This is because both the value of the \textit{synthetic} futures contract (an average value, i.e., an index) and the value of the financial futures contract (which is written on the index) should converge to the compute spot index value. Conversely, the physical access wedge should converge to a stable per-rental-period value as the maturity grows: $\lim_{T \to \infty} \overline{\Delta}^\Pi_t(t\rightarrow T) = \overline{\Delta}^{\Pi,\max} < \infty.$ Appendix~\ref{app:run-off-return-effect} formalizes these arguments with a simple \enquote{run-off} model. In this interpretation, the negative short-maturity returns are not evidence against a positive compute risk premium. They just reflect some inaccuracy when using \textit{synthetic} futures prices as a proxy for true futures price.

Across compute generations, the pattern in mean returns is not homogeneous. H100 performs especially well in this short sample, but much of this performance occurs around the late-2025 and early-2026 price increases associated with the agentic-AI demand wave. 

Turning to standard deviations, newer compute generations tend to have greater pricing uncertainty. A monotonic ordering from A100 to H100 to B200 holds at the one-month and 24-month maturities. B200 is also the most volatile at 12 months. 

The market beta estimates are mostly positive. They tend to rise with maturity, but the trend is weak and not monotonic. It is more visible for short and intermediate maturities. The long-maturity betas are less stable across compute generations.

\section{Conclusions}

Compute is a key input in the AI economy. Its spot price represents access to a scarce resource in AI production. Its forward price reflects uncertainty about model progress, data-center deployment, energy constraints and, more broadly, AI adoption. The imminent launch of compute futures is, therefore, an event worth studying: it will turn compute scarcity into a tradable risk to be priced. 

\begin{itemize}
\item We review the current compute service market, its indexation and the available (limited) data underlying the upcoming financialization of compute assets. 
\item We argue that, because compute is not storable, current spot prices cannot be directly translated into futures prices using  standard cash-and-carry no-arbitrage logic. In this sense, while it is sometimes described as \enquote{the new oil of 21st centrury,} compute is more like electricity than oil.
\item Term rental contracts provide a no-arbitrage benchmark for compute futures prices. Because of basis risk (due to physical access) forward prices - i.e., \textit{synthetic} futures prices in this article's terminology - are expected to be an upper bound on genuine compute futures prices. As financialization matures and compute intermediaries develop, basis risk will likely become more muted.
\item In a formal futures market, prices will reflect investors' expectation of spot compute prices at delivery adjusted by a risk premium. Should compute prices be negatively correlated with the marginal utility of the marginal investor, the compute risk premium would be positive. In the current economy, this will likely be the case for compute providers, i.e. the primary hedgers expected to drive futures prices upon launch of the new market.
\item Empirically, we construct the first panel of \textit{synthetic} futures curves, a cross section of GPU generations with prices organized by tenor. Preliminary empirical results using both hold-to-maturity (excess) returns and continuously-rolled constant-maturity (excess) returns are suggestive of a \textit{positive} risk premium. As emphasized in the previous bullet point, such a risk premium can be rationalized through hedging pressure exercised by compute-service providers. 
\end{itemize}

Our analysis aims at providing a framework for a new market and is, therefore, intentionally forward-looking. Needless to say, several important questions remain unaddressed: how to design \enquote{optimal} compute indexes pre-launch? How exactly the hedging pressure from providers and users will affect futures risk premia post-launch? What will the role of general financial market participants be as the market matures? Which factors will ultimately price compute risk? How will the wedge between forward and futures prices evolve as intermediaries and liquidity develop? How will the forward/futures curves discipline investment in chips, data centers, and energy? and so on. These issues - and many others - make compute asset-pricing an interesting area of research for financial economics in the age of AI.

\clearpage
\begin{singlespace}
\bibliography{computeAP_20260525,literature_review_candidates}

\begin{thebibliography}{27}
\providecommand{\natexlab}[1]{#1}
\expandafter\ifx\csname urlstyle\endcsname\relax
  \providecommand{\doi}[1]{doi:\discretionary{}{}{}#1}\else
  \providecommand{\doi}{doi:\discretionary{}{}{}\begingroup
  \urlstyle{rm}\Url}\fi

\bibitem[{Acemoglu(2024)}]{acemoglu2024simple}
Acemoglu, D. 2024.
\newblock The simple macroeconomics of {AI}.
\newblock NBER Working Paper 32487, National Bureau of Economic Research.
\newblock \doi{10.3386/w32487}.

\bibitem[{Akerlof(2020)}]{akerlof2020sins}
Akerlof, G.~A. 2020.
\newblock Sins of omission and the practice of economics.
\newblock \emph{Journal of Economic Literature} 58:405--18.

\bibitem[{Becker(1968)}]{becker1968}
Becker, G.~S. 1968.
\newblock Crime and punishment: An economic approach.
\newblock \emph{Journal of Political Economy} 76:169--217.

\bibitem[{Belt and Thomas(2026)}]{belt2026commodity}
Belt, A., and A.~Thomas. 2026.
\newblock Is {AI} computing power becoming a commodity?
\newblock Boston Consulting Group.
\newblock Published July 23, 2026.

\bibitem[{Bergemann and Deb(2025)}]{bergemann2025robust}
Bergemann, D., and R.~Deb. 2025.
\newblock Robust pricing for cloud computing.
\newblock Cowles Foundation Discussion Paper 2423, Cowles Foundation for
  Research in Economics, Yale University.

\bibitem[{Bessembinder and Lemmon(2002)}]{bessembinder2002equilibrium}
Bessembinder, H., and M.~L. Lemmon. 2002.
\newblock Equilibrium pricing and optimal hedging in electricity forward
  markets.
\newblock \emph{The Journal of Finance} 57:1347--82.

\bibitem[{Brynjolfsson, Rock, and Syverson(2017)}]{brynjolfsson2017paradox}
Brynjolfsson, E., D.~Rock, and C.~Syverson. 2017.
\newblock Artificial intelligence and the modern productivity paradox: A clash
  of expectations and statistics.
\newblock NBER Working Paper 24001, National Bureau of Economic Research.
\newblock \doi{10.3386/w24001}.

\bibitem[{Byrne, Corrado, and Sichel(2021)}]{byrne2021cloud}
Byrne, D., C.~Corrado, and D.~Sichel. 2021.
\newblock The rise of cloud computing: Minding your {P}s, {Q}s and {K}s.
\newblock In C.~Corrado, J.~Haskel, J.~Miranda, and D.~Sichel, eds.,
  \emph{Measuring and Accounting for Innovation in the Twenty-First Century},
  chap.~13. University of Chicago Press.

\bibitem[{Cochrane(2005)}]{cochrane2005asset}
Cochrane, J.~H. 2005.
\newblock \emph{Asset pricing}.
\newblock Revised ed. Princeton, NJ: Princeton University Press.
\newblock ISBN 9780691121376.

\bibitem[{{Commodity Futures Trading Commission}(2026)}]{cftc2026compute}
{Commodity Futures Trading Commission}. 2026.
\newblock Request for comment on the listing of compute derivatives contracts.
\newblock \emph{Federal Register} 91:54259--64.
\newblock August 21. FR Doc. 2026-17163.

\bibitem[{De~Roon, Nijman, and Veld(2000)}]{deroon2000hedging}
De~Roon, F.~A., T.~E. Nijman, and C.~Veld. 2000.
\newblock Hedging pressure effects in futures markets.
\newblock \emph{The Journal of Finance} 55:1437--56.

\bibitem[{Demirer et~al.(2025)Demirer, Fradkin, Tadelis, and
  Peng}]{demirer2025emerging}
Demirer, M., A.~Fradkin, N.~Tadelis, and S.~Peng. 2025.
\newblock The emerging market for intelligence: Pricing, supply, and demand for
  {LLM}s.
\newblock NBER Working Paper 34608, National Bureau of Economic Research.
\newblock \doi{10.3386/w34608}.

\bibitem[{Duffie and Dworczak(2021)}]{duffiedworczak2021}
Duffie, D., and P.~Dworczak. 2021.
\newblock Robust benchmark design.
\newblock \emph{Journal of Financial Economics} 142:775--802.

\bibitem[{Fama and French(1987)}]{fama1987commodity}
Fama, E.~F., and K.~R. French. 1987.
\newblock Commodity futures prices: Some evidence on forecast power, premiums,
  and the theory of storage.
\newblock \emph{The Journal of Business} 60:55--73.

\bibitem[{Friedman(2026)}]{friedman2026compute}
Friedman, D. 2026.
\newblock Compute is the commodity no one knows how to price.
\newblock Buy the Rumor; Sell the News.
\newblock Published February 4, 2026.

\bibitem[{He et~al.(2025)He, Manela, Ross, and von
  Wachter}]{he2025fundamentals}
He, S., A.~Manela, O.~Ross, and V.~von Wachter. 2025.
\newblock Fundamentals of perpetual futures.
\newblock SSRN Working Paper 4301150, SSRN.
\newblock \doi{10.2139/ssrn.4301150}.
\newblock Originally posted December 2022; last revised June 3, 2025.

\bibitem[{Kiyotaki and Moore(1997)}]{kiyotakimoore1997}
Kiyotaki, N., and J.~Moore. 1997.
\newblock Credit cycles.
\newblock \emph{Journal of Political Economy} 105:211--48.

\bibitem[{Korinek and Vipra(2024)}]{korinek2024concentrating}
Korinek, A., and J.~Vipra. 2024.
\newblock Concentrating intelligence: Scaling and market structure in
  artificial intelligence.
\newblock NBER Working Paper 33139, National Bureau of Economic Research.
\newblock \doi{10.3386/w33139}.

\bibitem[{Kumar and Seppi(1992)}]{kumarseppi1992}
Kumar, P., and D.~J. Seppi. 1992.
\newblock Futures manipulation with ``cash settlement''.
\newblock \emph{The Journal of Finance} 47:1485--502.

\bibitem[{Longstaff and Wang(2004)}]{longstaff2004electricity}
Longstaff, F.~A., and A.~W. Wang. 2004.
\newblock Electricity forward prices: A high-frequency empirical analysis.
\newblock \emph{The Journal of Finance} 59:1877--900.

\bibitem[{Morris and Shin(2002)}]{morrisshin2002}
Morris, S., and H.~S. Shin. 2002.
\newblock Social value of public information.
\newblock \emph{American Economic Review} 92:1521--34.

\bibitem[{Szymanowska et~al.(2014)Szymanowska, De~Roon, Nijman, and Van~den
  Goorbergh}]{szymanowska2014anatomy}
Szymanowska, M., F.~De~Roon, T.~Nijman, and R.~Van~den Goorbergh. 2014.
\newblock An anatomy of commodity futures risk premia.
\newblock \emph{The Journal of Finance} 69:453--82.

\bibitem[{Timmermann and Vulicevic(2026)}]{timmermann2026compute}
Timmermann, A., and L.~Vulicevic. 2026.
\newblock Compute, complexity, and the scaling laws of return predictability.
\newblock SSRN Working Paper 6105327, SSRN.
\newblock \doi{10.2139/ssrn.6105327}.

\bibitem[{Van~Nieuwerburgh(2026)}]{vannieuwerburgh2026financing}
Van~Nieuwerburgh, S. 2026.
\newblock Financing the {AI} buildout.
\newblock Working paper, Columbia Business School.

\bibitem[{Wachter and Wachter(2026)}]{wachter2026investment}
Wachter, J., and J.~Wachter. 2026.
\newblock What investment data implies about the {AI} transition.
\newblock NBER Working Paper 35290, National Bureau of Economic Research.
\newblock \doi{10.3386/w35290}.

\bibitem[{Wooldridge(2020)}]{wooldridge2020introductory}
Wooldridge, J.~M. 2020.
\newblock \emph{Introductory econometrics: A modern approach}.
\newblock 7 ed. Cengage Learning.
\newblock ISBN 9781337558860.

\bibitem[{Zhang(2022)}]{zhang2022}
Zhang, A.~L. 2022.
\newblock Competition and manipulation in derivative contract markets.
\newblock \emph{Journal of Financial Economics} 144:396--413.

\end{thebibliography}
\bibliographystyle{rfs}
\end{singlespace}
\clearpage
\appendix
\begin{center}
{\LARGE
\textbf{Appendix}
\\[12pt]
}
\end{center}
\renewcommand\thefigure{\thesection.\arabic{figure}}
\renewcommand\thetable{\thesection.\arabic{table}}
\renewcommand\theequation{\thesection.\arabic{equation}}

\section{Additional technical details}
\setcounter{figure}{0}
\setcounter{table}{0}
\setcounter{equation}{0}

\subsection{\textit{Synthetic} futures values with \textit{discounting}}
\label{app:synthetic-forward-discounting}

This subsection provides the time-discounted versions of the \textit{synthetic} futures definitions. Let \(PV_t(T)\) be the present value multiplier at \(t\) for one dollar paid at \(T\). Let \(APV_t(t\rightarrow T)\) be the annuity present value multiplier for one dollar paid over the rental interval \([t,T]\). In the continuous case:
\[
APV_t(t\rightarrow T)
=
\int_t^T PV_t(s)\,ds .
\]
The present value of the fixed rental strip is, therefore, 
\begin{eqnarray*}
APV_t(t\rightarrow T)\Pi_t(t\rightarrow T) 
=
\int_t^T PV_t(s) F^{\mathrm{syn}}_t(s)\,ds 
\end{eqnarray*}
and
\begin{eqnarray*}
\Pi_t(t\rightarrow T) 
=
\frac{\int_t^T PV_t(s) F^{\mathrm{syn}}_t(s)\,ds}{APV_t(t\rightarrow T)} = \frac{\int_t^T PV_t(s) F^{\mathrm{syn}}_t(s)\,ds}{\int_t^T PV_t(s)\,ds}. 
\end{eqnarray*}

Thus,

\begin{equation*}
F_t^{\mathrm{syn}}(T)
=
\frac{\frac{\partial}{\partial T}
\left[
APV_t(t\rightarrow T)\Pi_t(t\rightarrow T)
\right]}{PV_t(T)}.
\end{equation*}

The discrete \textit{synthetic} futures price for delivery month \(M\) is, therefore,
\begin{equation}
F_t^{\mathrm{syn},\mathrm{m}}(M) :=
\frac{
APV_t(t\rightarrow M)\Pi_t(t\rightarrow M)
- APV_t(t\rightarrow M-1)\Pi_t(t\rightarrow M-1)
}{
PV_t(M)
},
\label{eq:synthetic-forward-monthly-discounting}
\end{equation}
where \(APV_t(t\rightarrow M)\) is the discrete annuity multiplier
\[
APV_t(t\rightarrow M) =
\sum_{s\in \mathcal{M}(t,M)} PV_t(s),
\]
with \(\mathcal{M}(t,M)\) denoting the set of monthly payment dates in the rental strip from \(t\) to the delivery month \(M\). 

\subsection{Deriving the compute risk premium}
\label{app:risk-premium-sdf}

This subsection derives the compute risk premium expression used in Section~\ref{sec:risk-pricing}. We follow the canonical,  textbook treatment in, e.g., \citet{cochrane2005asset}.

A long futures position entered at date \(t\) for delivery at \(T\) has payoff \(S_T-F_t(T)\). Let \(M_{t,T}\) denote the stochastic discount factor from \(t\) to \(T\). The pricing condition is
\[
0 =
\mathbb{E}^{\mathbb{Q}}_t
\left[
S_T-F_t(T)
\right]
=
\mathbb{E}_t
\left[
M_{t,T}
\left(S_T-F_t(T)\right)
\right].
\]
Because \(F_t(T)\) is known at date \(t\), we have
\begin{eqnarray}\label{pri}
F_t(T) =
\frac{
\mathbb{E}_t[M_{t,T}S_T]
}{
\mathbb{E}_t[M_{t,T}]
}.
\end{eqnarray}
Re-expressing the numerator, we obtain
\[
\mathbb{E}_t[M_{t,T}S_T]
= \mathbb{E}_t[M_{t,T}]
\mathbb{E}_t[S_T]
+ \operatorname{Cov}_t(M_{t,T},S_T).
\]
Plugging this expression back into Eq. (\ref{pri}) gives
\[
F_t(T)
= \mathbb{E}_t[S_T]
+ \frac{
\operatorname{Cov}_t(M_{t,T},S_T)
}{
\mathbb{E}_t[M_{t,T}]
},
\]
and, under the sign convention in the main text,
\[
F_t(T)
= \mathbb{E}_t[S_T]
- \lambda_t(T).
\]
A long futures position has expected payoff
\begin{equation*}
\mathbb{E}_t
\left[
S_T-F_t(T)
\right]
= \lambda_t(T),
\end{equation*}
i.e., the compute risk premium.

\subsection{The relation between Silicon Data term rates and implied forward rates}\label{data1}

The exact proprietary transformation used by Silicon Data is not documented in the data available to us. We, therefore, treat the reported term rental strip and the reported synthetic forward rates as two observed data objects and infer the transformation that most likely maps one into the other.

Our replication exercise suggests that Silicon Data's forward rates are very close to the continuous no-discount version of the term-to-forward formula given in the main text (c.f. Eq. (\ref{eq:synthetic-forward-continuous-nodiscount})). In particular, let \(x\) denote tenor in months. Write \(\texttt{term\_SD}(t,x)\) for Silicon Data's reported \texttt{term\_rate} at quote date \(t\) and tenor \(x\), and \(\texttt{forward\_SD}(t,x)\) for the reported \texttt{forward\_rate}. On Silicon Data's fine tenor grid, whose spacing is \(\delta=0.25\) months, the reported \texttt{forward\_rate} is very close to the centered finite-difference derivative
\[
\begin{aligned}
\texttt{forward\_SD}(t,x)
&\approx
\frac{
(x+\delta)\texttt{term\_SD}(t,x+\delta)
-
(x-\delta)\texttt{term\_SD}(t,x-\delta)
}{
2\delta
} \\
&\approx
\frac{\partial}{\partial x}
\left[
x\texttt{term\_SD}(t,x)
\right].
\end{aligned}
\]
This is the continuous analogue of backing out the marginal rental price from the total value of a rental strip, while abstracting from discounting. The replication is not exact, which is unsurprising because Silicon Data may use smoothing, interpolation, day-count conventions, or other implementation choices that are not public. Numerically, however, the differences are small enough that we view the reported Silicon Data forward curve as consistent with a no-discount continuous construction.

\subsection{Implied forward rates: mapping tenors and expiration months}\label{data2}

In the article, \textit{synthetic} futures are indexed by \(t,M\) (quote date and delivery month) while the Silicon Data forward curve data is indexed by \(t,h\) (quote date and tenor), with $h$ on the discrete tenor grid \(\mathcal{H} = \{0,0.25,0.5,\ldots,36\}\). 

Thus, we need to map the delivery month \(M\) to the corresponding tenor \(h\) on the Silicon forward curve. We do so by rounding the difference between the delivery month and the quote date to the nearest tenor on the grid: \(h = \operatorname{round}_{\mathcal{H}}(\text{time-difference-in-months}(M, t))\). In other words, for quote date \(t\) and target delivery month \(M\), we define
\[
F^{\mathrm{syn}}_{t}(M)
:=
\texttt{forward\_SD}\left(t, \operatorname{round}_{\mathcal{H}}(\text{time-difference-in-months}(M, t))\right).
\]
The round function is defined as
\[
\operatorname{round}_{\mathcal{H}}(x)
:=
\arg\min_{h\in\mathcal{H}}
\left|
h-x
\right|,
\]
and \(\text{time-difference-in-months}(M, t)\) is the time difference between the end of month \(M\) and day \(t\), measured in months. For example, if \(t\) is August 15, 2025, and \(M\) is September 2025, then the time difference is 1.5 months.

\subsection{Equivalence between zero-tenor \textit{synthetic} futures prices and spot prices}
\label{app:zero-tenor-forward}

We evaluate the expression in Eq. \eqref{eq:synthetic-forward-continuous-nodiscount} for $T=t$. Equivalently, write
\[
F_t^{\mathrm{syn}}(t+h) =
\frac{\partial}{\partial h}
\left[
h\Pi_t(t\rightarrow t+h)
\right].
\]
By the product rule:
\[
F_t^{\mathrm{syn}}(t+h) =
\Pi_t(t\rightarrow t+h)
+ h\frac{\partial \Pi_t(t\rightarrow t+h)}{\partial h}.
\]
Evaluating at \(h=0\) gives
\[
F_t^{\mathrm{syn}}(t) =
\Pi_t(t)
+ \left.
h\frac{\partial \Pi_t(t\rightarrow t+h)}{\partial h}
\right|_{h=0}
= \Pi_t(t).
\]
A rental strip whose endpoint coincides with the quote date has zero length, thus its rental rate is the current spot (rental) rate.

In the discrete case, the first non-zero monthly \textit{synthetic} futures price is obtained by evaluating the strip formula in Eq. \eqref{eq:synthetic-forward-monthly-nodiscount} at month \(M=t+1\):
\[
F_t^{\mathrm{syn},\mathrm{m}}(t+1) =
\mathrm{time\_diff}(t+1,t)\Pi_t(t+1)
- \mathrm{time\_diff}(t,t)\Pi_t(t).
\]
Since \(\mathrm{time\_diff}(t,t)=0\) and \(\mathrm{time\_diff}(t+1,t)=1\) month, the expression reduces to
\(F_t^{\mathrm{syn},\mathrm{m}}(t+1) = \Pi_t(t+1)\), which is the spot (rental) rate for the upcoming month.

\subsection{Run-off of the physical access wedge}
\label{app:run-off-return-effect}

Our return construction uses \textit{synthetic} futures prices as stand-ins for true financial futures prices. 
The physical access wedge may, therefore, affect measured \textit{synthetic} returns, even if true financial futures follow the risk pricing logic in Section \ref{sec:risk-pricing}. Let \(h=T-t\) denote the time to delivery and write
\[
\overline{F}_t^{\mathrm{syn}}(T)
=
F_t(T)
+
\overline{\Delta}^F(h).
\]
Over a short holding period \(dt\), with the delivery date \(T\) fixed, the maturity falls from \(h\) to \(h-dt\). The \textit{synthetic} futures price change can be decomposed as
\[
\begin{aligned}
d\overline{F}_t^{\mathrm{syn}}(T)
&=
dF_t(T)
+
d\overline{\Delta}^F(h)
\\
&=
dF_t(T)
-
\frac{d\Delta^F(h)}{dh}dt .
\end{aligned}
\]
Equivalently, in return form,
\[
r_t^{\mathrm{syn}}(T)
\approx
r_t(T)
-
\frac{1}{F_t^{\mathrm{syn}}(T)}
\frac{d\overline{\Delta}^F(h)}{dh}dt .
\]
Thus, when \(\frac{d\overline{\Delta}^F(h)}{dh}>0\), the run-off of the physical access wedge lowers the measured synthetic return. The strength of this effect is governed by the derivative \(\frac{d\overline{\Delta}^F(h)}{dh}\). To illustrate, suppose the physical access wedge in rental-strip space takes the fractional form
\[
\overline{\Delta}^\Pi(h)
=
\widetilde{\Delta}^{\Pi}\frac{h}{h+a},
\qquad
\widetilde{\Delta}^{\Pi}>0,\quad a>0.
\]
This wedge is zero at \(h=0\), increases in \(h\) and converges to the finite per-period value \(\widetilde{\Delta}^{\Pi}\) as \(h\rightarrow\infty\). The wedge in the futures space is, instead,
\[
\begin{aligned}
\overline{\Delta}^F(h)
&=
\frac{d}{dh}
\left[
h\overline{\Delta}^\Pi(h)
\right]
\\
&=
\widetilde{\Delta}^{\Pi}
\frac{h(h+2a)}{(h+a)^2}.
\end{aligned}
\]
Its derivative is
\[
\frac{d\overline{\Delta}^F(h)}{dh}
=
\widetilde{\Delta}^{\Pi}
\frac{2a^2}{(h+a)^3}
>
0.
\]
This derivative is positive and decreasing in \(h\). Therefore, the run-off term is negative for \textit{synthetic} returns and relatively larger at short maturities.

\section{Additional calculations and empirical results}
\setcounter{figure}{0}
\setcounter{table}{0}
\setcounter{equation}{0}

\subsection{Compute service flow and capex to GDP}
\label{app:compute-scale-calculations}

This appendix documents the scale calculations used in the introduction. 

The U.S. nominal GDP at a seasonally-adjusted annual rate in 2026Q1 is \$31.86 trillion, according to FRED/BEA.

First, we calculate compute service flow as a gross rental-equivalent flow. According to Epoch AI, the installed AI accelerator stock is \(19{,}700{,}143\) H100-equivalent GPUs as of 2025Q4, after excluding chips reported for Huawei and Cambricon.\footnote{Epoch AI, ``Data on AI Chip Sales,'' 2026, accessed May 25, 2026, \href{https://epoch.ai/data/ai-chip-sales}{link}. We use the cumulative 2025Q4 stock and exclude Huawei and Cambricon. The remaining H100-equivalent stock is mostly Nvidia, with AMD, Amazon, and Google chip lines translated into Nvidia H100-equivalent units.} According to Silicon Data, on April 18, 2026, the H100 neocloud rental index was \(\$2.50\) per GPU-hour and the H100 hyperscaler rental index was \(\$7.43\) per GPU-hour.\footnote{Silicon Data, ``Silicon H100 Rental Index,'' 2026, accessed May 25, 2026, \href{https://portal.silicondata.com/gpu-index-chart}{link}.} Hyper-scalers currently account for most cloud compute capacity. We do not have a precise market-share weight between hyper-scaler and neo-cloud pricing. The gross annual rental-equivalent service flow is installed units times rental price times hours per year, assuming full utilization. Using the neo-cloud rental price, we have
\[
19{.}70 \text{ mil.\ H100e} \times \$2.50 \text{ per GPU-hour} \times 8{,}760 \text{ hours per year}
= \$431.433 \text{ billion}.
\]
Using the hyper-scaler rental price, we have
\[
19{.}70 \text{ mil.\ H100e} \times \$7.43 \text{ per GPU-hour} \times 8{,}760 \text{ hours per year}
= \$1{,}282.219 \text{ billion}.
\]
Dividing by U.S. nominal GDP, we have
\[
\frac{\$431.433 \text{ billion}}{\$31{,}856.257 \text{ billion}}=1.35\%,
\]
based on neo-clouds rental prices and 
\[
\frac{\$1{,}282.219 \text{ billion}}{\$31{,}856.257 \text{ billion}}=4.03\%,
\]
based on hyper-scalers rental prices. This calculation should not be interpreted as provider revenues or GDP value added. It is a full-utilization rental-equivalent flow designed to measure the economic scale of the installed compute stock at prevailing rental prices.

Second, Bloomberg reports that the four largest U.S. hyper-scalers plan to spend as much as \$725 billion on capital expenditures in 2026, primarily on AI data-center equipment.\footnote{Vlad Savov, ``US big tech ratchets up AI spending past \$700 billion this year,'' \emph{Bloomberg}, April 30, 2026, \href{https://www.bloomberg.com/news/articles/2026-04-30/us-big-tech-ratchets-up-ai-spending-past-700-billion-this-year}{link}.} The component figures are approximately \(\$190\) billion for Microsoft, \(\$200\) billion for Amazon, \(\$180\)--\(\$190\) billion for Alphabet and \(\$125\)--\(\$145\) billion for Meta. The corresponding GDP share is
\[
\frac{\$725 \text{ billion}}{\$31{,}856.257 \text{ billion}}=2.28\%.
\]
This is a broad measure of global capital expenditure by U.S.-based firms, not a literal measure of U.S.-located investment. Also, not every dollar is AI-only compute infrastructure. It is, nevertheless, a relevant public capex scale for the firms currently building the dominant AI infrastructure platforms.

\subsection{Full sample rolling constant-maturity returns}
\label{app:full-sample-returns}

\begin{figure}[!t]
\centering
\includegraphics[width=\linewidth]{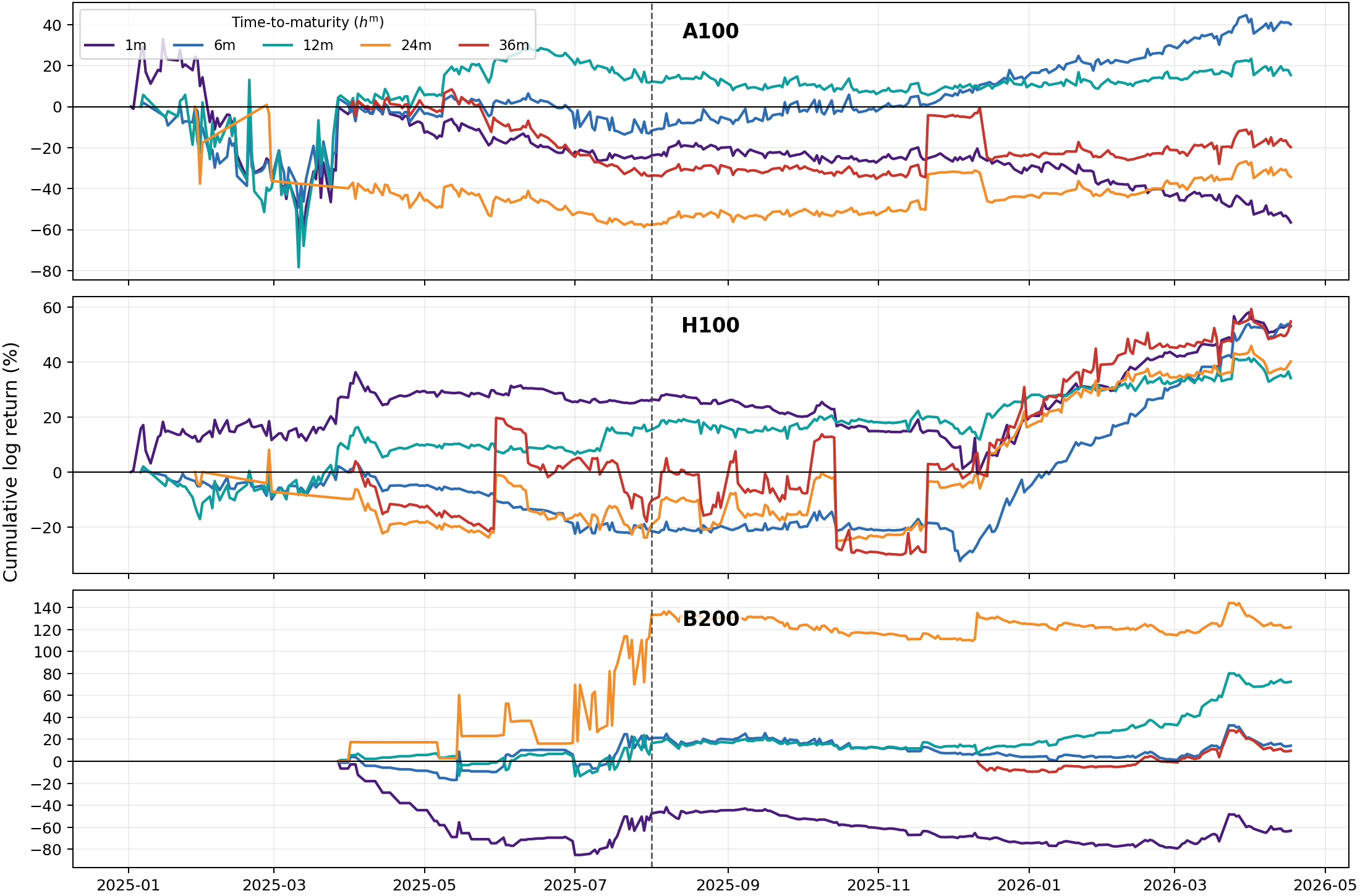}
\caption{\textbf{Full-sample rolling constant-maturity cumulative returns.} Cumulative log daily returns of rolling constant-maturity \textit{synthetic} futures positions using the longest available history for each GPU-maturity pair. The panels refer to vintage A100, H100 and B200, respectively. The dashed vertical line marks August 1, 2025, i.e., the start date of Fig.~\ref{fig:constant-maturity-cumulative} in the main text. It provides the time from which we are more confident in the quality of the Silicon Data forward curves.}\label{fig:full-sample-returns}
\end{figure}

Fig.~\ref{fig:full-sample-returns} reports rolling constant-maturity return paths using the longest available history for each GPU-maturity pair. The dashed vertical line marks August 1, 2025, i.e., the start date used in the aligned sample in the main text.

\pagebreak

\section{Comments on the Listing of Compute Derivatives Contracts}\label{comments}

We appreciate the opportunity to comment on the Commodity Futures Trading Commission's Request for Comment on the Listing of Compute Derivatives Contracts.\footnote{These comments were submitted in response to the Commodity Futures Trading Commission's \enquote{Request for Comment on the Listing of Compute Derivatives Contracts}, published August 21, 2026, at 91 FR 54259 (RIN 3038-AF77; Docket No. CFTC-2026-1850).} The Commission cites an earlier version of this paper. For conciseness, we summarize our principal conclusions below.

\begin{enumerate}[leftmargin=*,itemsep=0.5em]
\item \textbf{Technological obsolescence.}
Individual hardware configurations becoming obsolete need not prevent a viable futures market. Contracts can reference specific configurations, with contracts on new configurations introduced as they become available. For each configuration, the term structure can reflect expected obsolescence.

\item \textbf{Service heterogeneity.}
Even for a given hardware configuration, the economic value of a GPU-hour can differ across providers because of, e.g., form factor, interconnectivity, and network specifications. A clearly specified reference, or \enquote{par}, service may define the underlying, with transparent price adjustments for deviations from the \enquote{par} characteristics.

\item \textbf{OTC transactions and reproducibility.}
A settlement index should be simple and reproducible from verifiable transactions. A trade-reporting infrastructure that records transactions while protecting privacy can support this objective. Where transaction liquidity is insufficient, firm, binding quotes from a contributor panel could supplement the inputs. Nonbinding indicative quotes should be identified separately and their role in validation and weighting made transparent. 

\item \textbf{Manipulation resistance.}
In a simple model, below, we suggest that regulators could use unusually large net, settlement-linked futures exposures as a screening signal for closer scrutiny. The model’s exposure threshold should, however, only serve as an illustrative benchmark—not as a validated all-encompassing rule for detecting manipulation. The model also highlights index-design choices that can reduce incentives to distort settlement, including broad and transparent survey coverage, carefully chosen contributor weights, trimming extreme observations, and averaging prices over the settlement period. These safeguards involve trade-offs: weighting affects both manipulation incentives and representativeness, while averaging may reduce manipulation sensitivity but increase basis risk for users exposed near expiry. Because the suggested threshold rests on stylized assumptions about price distortion, economic costs, and exposure, it should inform surveillance and index-design due diligence. It should not be viewed as replacing documented evidence of misconduct or a well-calibrated regulatory standard.

\end{enumerate}

We emphasize that our objective is not advocacy. Our contribution and interests are technical. While we do not believe that compute's distinctive features are an insurmountable obstacle to a viable futures market, compliance with the IOSCO principles requires care and due diligence.

Subsection~\ref{sec:model} develops a simple model of manipulation incentives and its implications for regulation and index construction. Subsection~\ref{sec:representativeness} considers benchmark representativeness, the usefulness of a contract covering a defined market segment, and the risks of over-interpreting its price signals.

\subsection{A simple model of compute index manipulation: implications for regulation and index construction}
\label{sec:model}
We study the potential manipulation of the compute rental index for gains in the futures market, as well as its deterrence, by analyzing the incentives and costs of such manipulation using a rational-choice approach to crime and deterrence \citep{becker1968}. The model compares the gain on an index-linked financial position with the economic cost of distorting the index and the expected penalty. Under simplifying assumptions, the model provides an exposure threshold above which distortion is economically profitable, informing regulatory guidelines for monitoring position sizes. It also identifies how various index construction considerations can reduce manipulation incentives, informing the design of manipulation-resistant settlement indices.

Suppose provider $i$ distorts its reported rental price by $x_i$ dollars per GPU-hour for one day. Its profit from such distortion is
\begin{equation}
\label{eq:manipulation-profit}
\begin{array}{@{}c@{\;}c@{\;}c@{\;}c@{\;}c@{\;}c@{\;}c@{}}
\Pi_i(x_i) & = & q_i\Delta_Y(x_i) & - & C_i(x_i) & - & \text{expected penalty}_{i} \\[2pt]
\downarrow & & \downarrow & & \downarrow & & \downarrow \\[2pt]
\substack{\text{provider }i\text{'s profit}\\\text{from price distortion }x_i}
& & \substack{\text{benefit of index distortion}\\=\text{contract exposure}\\\times\text{index distortion}}
& & \substack{\text{economic cost}\\\text{of price distortion}}
& & \substack{\text{expected penalty}\\\text{for provider }i}
\end{array}
\end{equation}
Here, $\Pi_i(x_i)$ is provider $i$'s incremental profit. When $\Pi_i(x_i)>0$, the provider has a net incentive to conduct manipulation. The quantity $q_i$ is provider $i$'s effective settlement-linked financial exposure, positive for a long position and negative for a short; in an application, relevant offsets and related positions should be considered. (Exposure is measured in GPU-hours, thus one unit of $q_i$ represents one GPU-hour of exposure to the index. Under the proposed CME contract specification, one contract represents 730 GPU-hours or, in other words, one GPU-month. The conversion is $Q_i=q_i/730$.) The quantity $\Delta_Y(x_i)$ is the change in the final settlement value induced by provider $i$'s distortion of $x_i$. $C_i(x_i)$ is the economic cost of the one-day reporting distortion, e.g., from charging a suboptimal price in the compute rental market. The final term is the expected penalty due to index manipulation, which is driven by the probability of sanction and punishment conditional on sanction.

Without specifying the details of index construction, Eq.~\eqref{eq:manipulation-profit} yields the following implications for monitoring financial positions and improving manipulation resistance.
\begin{itemize}[leftmargin=*,itemsep=0.5em]
\item \textbf{Financial exposure.} Larger exposures $|q_i|$ increase financial benefit and favor distortions. Monitoring unusually large exposures is therefore a relevant consideration. The more detailed quantitative analysis below provides a threshold for the magnitude of $q_i$ above which manipulation is profitable.

\item \textbf{Sensitivity of the settlement index.} The financial benefit depends on how strongly an individual reporting distortion changes settlement, as described by $\Delta_Y(x_i)$. A general design objective is to limit the magnitude of $\left|\frac{\partial\Delta_Y(x_i)}{\partial x_i}\right|$ (i.e., the sensitivity of the index to individual price distortions), while preserving the index's ability to represent rental prices. This consideration is relevant to index design choices such as volume weighting, averaging over the settlement month, contributor weight caps, and trimming extreme price inputs \citep{duffiedworczak2021}.

\item \textbf{Economic cost of distortion.} The cost $C_i(x_i)$ depends on the industrial organization of the compute rental spot market and the obligations associated with reported prices. Competition, customer switching, capacity utilization, and execution commitments determine whether manipulation entails forgone revenue, unfavorable transactions, or other economic penalties. 

\item \textbf{Legal and institutional environment.} The expected monetary punishment depends on the probability of being sanctioned and the sanction, conditional on enforcement. Auditing, surveillance, and legal jurisdiction shape these terms. We take the typical institutional framework for deterring manipulation to be applicable to the compute market. In this sense, the last term in Eq. (\ref{eq:manipulation-profit}) is not the focus of our simple model, which examines compute-specific issues. Having made this point, legal and institutional deterrents are important.
\end{itemize}

\subsubsection{Adding details to the model}
Next, we introduce simplifying assumptions in order to derive sharper quantitative implications. We follow \citet{duffiedworczak2021}, \citet{kumarseppi1992}, and \citet{zhang2022}.

\begin{enumerate}
\item Assume that settlement is the average of daily observations on the index over month $t$:
\begin{equation}
\label{eq:monthly-settlement}
Y_t=\frac{1}{|\mathcal D_t|}\sum_{\tau\in\mathcal D_t}Y_\tau,
\qquad
Y_\tau=\sum_{i=1}^{N}W_i P_{i\tau}.
\end{equation}
Here $\mathcal D_t$ is the set of days included in settlement for month $t$ and $|\mathcal D_t|$ denotes the number of days in that set. $Y_t$ is the monthly settlement average, $Y_\tau$ is the daily index, $P_{i\tau}$ is constituent $i$'s reported rental price on day $\tau$, and $W_i>0$ is its weight, with $\sum_{i=1}^{N}W_i=1$. Assume homogeneous GPU services, with $N$ providers contributing to the index. Two examples of how to set the weights are
\begin{equation}
\label{eq:index-weights}
W_i=\frac{V_i}{V_{\mathrm{AGG}}}
\quad\text{(volume weighting)} \qquad\text{or}\qquad
W_i=\frac{1}{N}
\quad\text{(equal weighting)},
\end{equation}
where $V_i$ is the number of continuously rented GPUs supplied by provider $i$ and $V_{\mathrm{AGG}}=\sum_{i=1}^N V_i$ is the total stock of GPUs covered by the index survey sample. Quantities and weights are held fixed. Thus, a price distortion $x_i$ on one specific day changes the settlement value by $\Delta_Y(x_i)=W_i x_i/|\mathcal D_t|$.

\item Assume a linear economic cost, i.e., $C_i(x_i)=24\gamma V_i|x_i|$, where $\gamma>0$ is the economic cost per affected GPU-hour per dollar of price distortion, and assume that the distortion applies throughout the day for 24 hours to all $V_i$ GPUs supplied by provider $i$. The case $\gamma=1$ corresponds to sacrificing one dollar on every affected GPU-hour for a one-dollar price reduction, with rental quantities held fixed. Values $\gamma<1$ describe a cheaper distortion technology. This cost assumption is imposed for the direction of distortion under consideration; upward and downward distortions need not have the same cost in practice.

\item As mentioned, we ignore the expected penalty term to focus on the providers' main trade-off.
\end{enumerate}

Consider a price distortion in the direction favored by the financial position. The marginal profit per additional dollar of distortion in the reported price for one day is
\begin{equation}
\label{eq:marginal-profit}
\pi_i\equiv\frac{\partial\Pi_i}{\partial |x_i|}
=\frac{|q_i|W_i}{|\mathcal D_t|}-24\gamma V_i.
\end{equation}
The pre-penalty threshold for positive marginal gains from the specified manipulation is therefore
\begin{equation}
\label{eq:exposure-cutoff}
q_i^{\mathrm M}=\frac{24|\mathcal D_t|\gamma V_i}{W_i},
\qquad
\pi_i>0\quad\Longleftrightarrow\quad |q_i|>q_i^{\mathrm M},
\end{equation}
where the last condition refers to distortion in the direction favored by the position.

Express exposure in standardized GPU-months as $Q_i=q_i/730$, with one contract-month defined as 730 GPU-hours (a standardized contract unit, not the exact number of hours in every calendar month). This equals the signed number of contracts for a futures contract with a 730 GPU-hour multiplier. The exact threshold in these units is $Q_i^{\mathrm M}=(24|\mathcal D_t|/730)\gamma V_i/W_i$. For the simple benchmark comparisons below, take settlement observations to cover calendar days and use $24|\mathcal D_t|\approx730$. Under this approximation, the threshold simplifies to $Q_i^{\mathrm M}=\gamma V_i/W_i$. Business-day sampling would suggest retaining the factor $24|\mathcal D_t|/730$.\footnote{The proposed NYMEX rules in Submission 26-370, August 11, 2026, specify a 730 GPU-hour contract unit and business-day averaging for the H100 and B200 contracts. See \href{https://www.cftc.gov/filings/ptc/ptc08112615405.pdf}{Exhibit A, Chapters 1045 and 1047}. The linear index and cost assumptions here are illustrative, not a description of those indices' full methodologies.}

This is a pre-penalty marginal-incentive threshold for the particular unilateral distortion and cost technology specified above. It is not, by itself, a threshold for the optimal distortion or a general measure of manipulation risk. A provider with net futures exposure $|Q_i|\leq Q_i^{\mathrm M}$ has nonpositive marginal gains from increasing that distortion under the model, before expected penalties. Exposure above the threshold makes the marginal gain positive. This conclusion conditions on an existing financial position and omits the cost of acquiring, financing, or unwinding it.

\subsubsection{Discussion}

\noindent\textbf{Total survey size as a monitoring threshold.}

Under the volume-weighting scheme, $W_i=V_i/V_{\mathrm{AGG}}$. This scheme would lead to a simple expression:
\begin{equation}
\label{eq:volume-weighted-threshold}
Q_i^{\mathrm M}=\gamma V_{\mathrm{AGG}}.
\end{equation}
Individual volume $V_i$ cancels out, so the same absolute exposure threshold applies to small and large providers. This cancellation is a consequence of the assumed linear cost, which scales with all $V_i$ GPUs supplied by the provider, and of volume weights proportional to $V_i$. It need not hold if the price distortion affects only a subset of capacity, if reported prices are nonbinding, or if the economic cost does not scale proportionately with supplied volume.

A larger total stock of GPUs covered by the index survey raises the exposure threshold for profitable manipulation. Under this assumption, therefore, broader survey coverage deters manipulation regardless of how the covered stock is distributed across providers.

As a numerical example, suppose the index provider's survey represents a total stock of $V_{\mathrm{AGG}}=10{,}000$ GPUs across all surveyed providers and $\gamma=1$. Under the calendar-day approximation above, any single provider with \textit{net} futures exposure exceeding 10,000 GPU-months (futures positions of $10{,}000\times730$ GPU-hours) has a positive marginal incentive to manipulate the index unilaterally by changing its reported rental price in the direction favored by its position.

In this example, 10,000 GPU-months is a model-based candidate exposure level for enhanced scrutiny, not a validated regulatory safe harbor or a stand-alone surveillance threshold. The appropriate monitoring level should depend on the index methodology, the quantity actually affected by a provider's price change, the provider's effective settlement-linked exposure (including relevant offsets), and the plausibility of the assumed cost technology. 

A lower economic cost of distortion ($\gamma<1$) reduces the model's marginal-incentive threshold.

\noindent\textbf{Distinguishing manipulation-prone positions from ordinary hedging positions.}

Next, we compare the manipulation threshold with the position needed to hedge a provider's rental revenue. The goal is to determine the extent to which profitable manipulation requires exposure beyond ordinary hedging needs, and how the distinction between manipulation and legitimate hedging needs depends on the provider's index weight and distortion cost.

For a stylized benchmark, suppose a common price movement changes the provider's rental price and the settlement index equally, and the provider's revenue exposure is $730V_i\Delta p$. A futures position of magnitude $Q_i^{\mathrm H}$ offsets this change when $730Q_i^{\mathrm H}\Delta p=730V_i\Delta p$. Thus, given this benchmark (and the calendar-day approximation), the hedge position and its relation to the manipulation threshold are
\begin{equation}
\label{eq:hedge-comparison}
Q_i^{\mathrm H}=V_i,
\qquad
\frac{Q_i^{\mathrm M}}{Q_i^{\mathrm H}}=\frac{\gamma}{W_i}.
\end{equation}
Under this stylized hedge benchmark, when $\gamma=1$ and volume weighting gives a provider a $W_i=V_i/V_{\mathrm{AGG}}=10\%$ index weight, the marginal-manipulation threshold is ten times the benchmark hedge. (This gap may be informative for monitoring, but it is not a universal separation between manipulation-prone positions and legitimate hedges: actual hedge needs depend on revenue sensitivity, contract-index covariance, service mix, and any basis risk.) If the index is highly concentrated ($W_i$ close to one), or if distortion is relatively cheap ($\gamma<W_i$), the model's threshold can approach or fall below the benchmark hedge. Anticipated manipulation profits can themselves encourage over-hedging, as analyzed by \citet{zhang2022}. Suppose we switch to equal weighting. Then $Q_i^{\mathrm M}/Q_i^{\mathrm H}=\gamma N$, so the ratio of the surveillance threshold to the ordinary hedging position is constant across small and large providers, thereby offering an equal proportional buffer for surveillance monitoring. Holding $\gamma$ fixed, surveying more providers (a larger $N$) increases this surveillance buffer ratio.

\noindent\textbf{Index construction method: volume weighting versus equal weighting.}

As a general principle within this simple model, volume weighting makes larger providers more susceptible to manipulation, whereas equal weighting makes smaller providers more susceptible. The choice between the two, or a weighting method that lies between these two regimes, depends on the following trade-offs.

If monitoring centers on the absolute size (number of contracts) of the manipulation threshold $Q_i^{\mathrm M}$, volume weighting provides a common surveillance level across providers, as shown in Eq.~\eqref{eq:volume-weighted-threshold}. This result depends on the assumption that the economic cost of distortion ($C_i(x_i)$) also scales with the provider's size ($V_i$). When this assumption does not hold, as may be the case for nonbinding posted prices rather than transaction prices, the benefit of volume weighting is less clear, and a weighting method closer to equal weighting may be considered.

If monitoring instead focuses on distinguishing manipulation-prone positions from ordinary hedging positions, the relevant measure is the ratio $Q_i^{\mathrm M}/Q_i^{\mathrm H}$. Under volume weighting, if one provider accounts for most of the surveyed volume, for example $W_i=V_i/V_{\mathrm{AGG}}=75\%$, its manipulation threshold is only $4/3$ of its ordinary hedge when $\gamma=1$. This leaves little room to distinguish manipulation-prone positions from ordinary hedging positions. Capping very large contributor weights may therefore be useful, moving the weighting toward a more equal allocation.\footnote{Regardless of weighting, such concentration would raise a question about the index's representativeness. More on this issue below.} Equal weighting sets $W_i=1/N$, so Eq.~\eqref{eq:hedge-comparison} gives the same ratio $\gamma N$ for every provider. Moving toward equal weighting therefore widens the proportional surveillance buffer for larger providers while narrowing it for smaller providers. This favors more equal weighting when the concern is that a large provider's manipulation threshold lies too close to its ordinary hedge. The trade-off is a lower absolute threshold for smaller providers.

\noindent\textbf{Index construction method: trimming extreme price inputs.}

Suppose a trader already has an unusually large exposure that exceeds the threshold for marginally profitable manipulation ($|Q_i|>Q_i^{\mathrm M}$). Under the linear benefit and cost assumptions, its total profit before penalty, $\pi_i|x_i|$, continues to increase with larger distortions $|x_i|$. Within this linear model, the only remaining potential deterrent to ever-larger distortions is the expected penalty: larger distortions may become more conspicuous, increasing the ex-ante probability of sanction and hence the expected penalty.

Trimming extreme price inputs in index construction can provide another channel for deterring ever-increasing manipulation. Trimming would therefore exclude observations in specified lower and upper tails of the price distribution when calculating the index. Once a price change is excluded (and the retained observations and their weights are unchanged), further distortions have no marginal effect on settlement, i.e., $\partial\Delta_Y(x_i)/\partial x_i=0$. Thus, trimming is a reasonable index design choice for limiting manipulation gains from extreme price inputs.

A provider may therefore recognize that trimming can bound its gains, even when $|Q_i|>Q_i^{\mathrm M}$, if the rule bounds the attainable settlement change for that position. This conclusion is conditional: a price input must remain outside the retained set, and the retained observations and weights must not change in response to the distortion. In practice, rank changes, cutoff effects, and changes in the set of retained contributors can make the settlement impact nonzero. If the maximum attainable gain, net of expected penalties, is negative, the manipulation incentive would dissipate.

\noindent\textbf{Index construction method: averaging over the settlement month.}

Averaging over the settlement month dilutes the effect of a one-day distortion. Under the model's assumptions, replacing the monthly average with a single-day fixing lowers the threshold by a factor of $| \mathcal D_t|\approx 30$. The dilution is weaker if a distortion can be sustained over multiple settlement days. The trade-off is that an average responds less to prices near expiry and may increase \textit{basis risk} for users whose exposure is concentrated at that date rather than spread over the settlement month.

\subsection[Some observations regarding representativeness and transparency]{\fontsize{16.5}{20}\selectfont Considerations regarding representativeness\\and transparency}
\label{sec:representativeness}

Much of compute pricing is currently OTC. If the objective were to obtain a truly representative index which is easily replicable by market participants and auditors, the relevant transactions would have to be readily available. A trade-reporting infrastructure that records transactions while protecting privacy can support this objective. Where transaction liquidity is insufficient, firm, binding quotes from a contributor panel could supplement the inputs; nonbinding indicative quotes should be distinguished and their use in validation and weighting made transparent. 

We argue, however, that the first batch of compute futures contracts need \textit{not} represent the entire compute economy. Given the size of the compute market and the substantial uncertainty surrounding it, financializing a clearly-defined small segment of the compute universe can have substantial economic value by serving the risk-management needs of users and providers in that segment and beyond (through cross-hedging). In other words, contracts need clearly-defined economic exposures and credible settlement benchmarks. Their scope may, however, be more limited than representing the entire compute ecosystem, at least initially.

We further argue that interpreting the first compute futures contracts as representative of a broader market than they actually cover could have unintended consequences. The following channels are conditional risks: they require market participants to treat a narrow futures benchmark as a broad signal and lenders/investors to respond to it. If futures prices are mistakenly treated as a general guide to AI infrastructure investment and financing, they could distort credit decisions and amplify the credit cycle. Two channels from the existing literature help explain this concern. First, a coordination channel can arise when participants care about others' responses as well as fundamentals \citep{morrisshin2002}. Because AI infrastructure financing relies heavily on syndicated loans, private credit, and large debt packages, lenders may look for any forward-looking price anchor. If market participants treat compute futures as a general \enquote{health index} for the entire AI sector, compute futures may lead to a coordination trap. Lenders might keep funding marginal projects simply because they think other lenders are looking at the same futures signal and doing the same—until sentiment shifts, leading to a sudden, coordinated freeze. Second, a distinct collateral channel could lead to disruptions \citep{kiyotakimoore1997}. AI infrastructure is exceptionally capital-intensive, and much of it is debt-financed using the underlying physical assets (hardware, facilities, future cash-flow contracts) as collateral. If a drop in compute futures prices signals falling future utilization or margins, lenders may immediately mark down the value of existing infrastructure collateral. This may trigger margin calls, tighter borrowing limits, forced liquidations, and/or distressed asset sales. Those fire sales then would depress actual asset values further, creating a vicious feedback loop between the financial market and physical capital expenditure. We note, once more, that these channels represent theoretical (tail) risk mechanisms, rather than the likely effect of a well-functioning compute futures market.

In sum, we support transparency enabling market participants to understand precisely which market segment the compute index surveys and measures. Identifying an index as covering, e.g., the neocloud segment is helpful. Specifying which portion of that segment the index covers is necessary. This includes which providers are surveyed, which parts of their rental business enter the index, and how the surveyed observations are used to measure prices in that segment. 

A revealing analogy is a hotel room rate index within the broader real estate economy. While real estate largely consists of heterogeneous assets and long-term, private contracts, hotels feature shorter commitments and frequent repricing. Consequently, a room rate index for a specific hotel segment provides valuable price signals without tracking the entire real estate market. Similarly, a compute rental index may initially focus on a distinct segment, such as the on-demand neocloud market—a sector that is economically substantial in its own right. While some connection to the broader compute market is expected, strong price correlation should not be presumed. Just as hotel room rates relate to the wider real estate market, significant structural differences between the two remain.

\end{document}